\documentclass[prd, floatfix, amsfonts, amssymb, amsmath, reprint, showkeys, nofootinbib, twoside,superscriptaddress]{revtex4-2}
\usepackage{amsmath,color}
\usepackage{comment}

\usepackage{gensymb}
\usepackage{placeins}
\usepackage[dvipsnames, usenames]{xcolor}
\definecolor{linkcolor}{rgb}{0.0,0.3,0.5}
\usepackage[unicode, colorlinks=true, linkcolor=linkcolor, citecolor=linkcolor, filecolor=linkcolor, urlcolor=linkcolor, linktocpage, breaklinks]{hyperref}
\usepackage[all]{hypcap}
\usepackage[T1]{fontenc}
\usepackage[utf8]{inputenc}
\usepackage{mathrsfs}
\usepackage[usenames,dvipsnames]{xcolor}
\hypersetup{colorlinks=true,citecolor=romared,linkcolor=romared,urlcolor=romared}

\definecolor{romared}{RGB}{142,0,28}

\usepackage[normalem]{ulem}
\newcommand{\red}{\color{red}}

\usepackage{dcolumn}
\usepackage{hyphenat}
\usepackage{graphicx}
\usepackage{ulem}
\usepackage{aas_macros}

\usepackage{orcidlink}
\newcommand{\orcid}[1]{\href{https://orcid.org/#1}{\includegraphics[width=10pt]{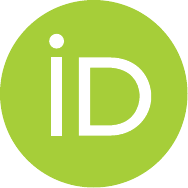}}}

\newcommand{\MinMass}{$1\, M_\odot$}
\newcommand{\MaxMass}{$2\, M_\odot$}

\begin{document}

\title{Constraining properties of asymmetric dark matter candidates from gravitational-wave observations}
\author{Divya Singh \orcid{0000-0001-9675-4584}}
\email{dus960@psu.edu}
\affiliation{Institute for Gravitation and the Cosmos, Department of Physics, Pennsylvania State University, University Park, PA, 16802, USA}

\author{Anuradha Gupta \orcid{0000-0002-5441-9013}}
\affiliation{Department of Physics and Astronomy, The University of Mississippi, University MS 38677, USA}

\author{Emanuele Berti \orcid{0000-0003-0751-5130}}
\affiliation{Department of Physics and Astronomy, Johns Hopkins University, 3400 N. Charles Street, Baltimore, Maryland, 21218, USA}

\author{Sanjay Reddy \orcid{0000-0003-3678-6933}}
\affiliation{Institute for Nuclear Theory, University of Washington, Seattle, WA USA}

\author{B. S. Sathyaprakash \orcid{0000-0003-3845-7586}}
\affiliation{Institute for Gravitation and the Cosmos, Department of Physics, Pennsylvania State University, University Park, PA, 16802, USA}
\affiliation{Department of Astronomy \& Astrophysics, Pennsylvania State University, University Park, PA, 16802, USA}
\affiliation{School of Physics and Astronomy, Cardiff University, Cardiff, UK, CF24 3AA}

\begin{abstract}
  The accumulation of certain types of dark matter particles in neutron star cores due to accretion over long timescales can lead to the formation of a mini black hole. In this scenario, the neutron star is destabilized and implodes to form a black hole without significantly increasing its mass.  When this process occurs in neutron stars in coalescing binaries, one or both stars might be converted to a black hole before they merge. Thus, in the mass range of $\sim \mbox{1--2}\, M_\odot,$ the Universe might contain three distinct populations of compact binaries: one containing only neutron stars, the second population of only black holes, and a third, mixed population consisting of a neutron star and a black hole. However, it is unlikely to have a mixed population as the various timescales allow for both neutron stars to remain or collapse within a short timescale. In this paper, we explore the capability of future gravitational-wave detector networks, including upgrades of Advanced LIGO and Virgo, and new facilities such as the Cosmic Explorer and Einstein Telescope (XG network), to discriminate between different populations by measuring the effective tidal deformability of the binary, which is zero for binary black holes but nonzero for binary neutron stars. Furthermore, we show that observing the relative abundances of the different populations can be used to infer the timescale for neutron stars to implode into black holes, and in turn, provide constraints on the particle nature of dark matter.  The XG network will infer the implosion timescale to within an accuracy of 0.01 Gyr at 90\% credible interval and determine the dark matter mass and interaction cross section to within a factor of 2 GeV and 10 cm$^{-2}$, respectively.

\end{abstract}

\keywords{Gravitational Waves; Astrophysics, Black Holes, Star Clusters, Dark Matter}
\maketitle

\section{Introduction and background}
\label{sec:intro}

The origin and properties of dark matter (DM) have been long-standing problems in fundamental physics and cosmology. Astronomical observations have increasingly provided evidence for a non-baryonic component of matter that either does not interact electromagnetically with baryons or has a negligibly small interaction cross-section. Consequently, the presence of DM is inferred due to its gravitational effect on baryonic matter. Laboratory experiments to detect DM particles from their weak interaction with baryons  have so far produced null results, as have the observations of decay products that would result from the annihilation of certain types of DM particles. Although there are a few plausible DM candidates in the Standard Model, theoretical insight into what they might be in theories beyond the Standard Model is plentiful and not very constraining. Currently, there is an effort to look for DM over sixty orders of magnitude in mass, with candidates ranging from wave-like~\cite{Marsh:2021lqg} and particle DM~\cite{Bertone:2004pz} to macroscopic objects such as primordial black holes (BHs)~\cite{Carr:2020xqk}.

\begin{figure*}
    \centering
    \includegraphics[width=0.95\textwidth]{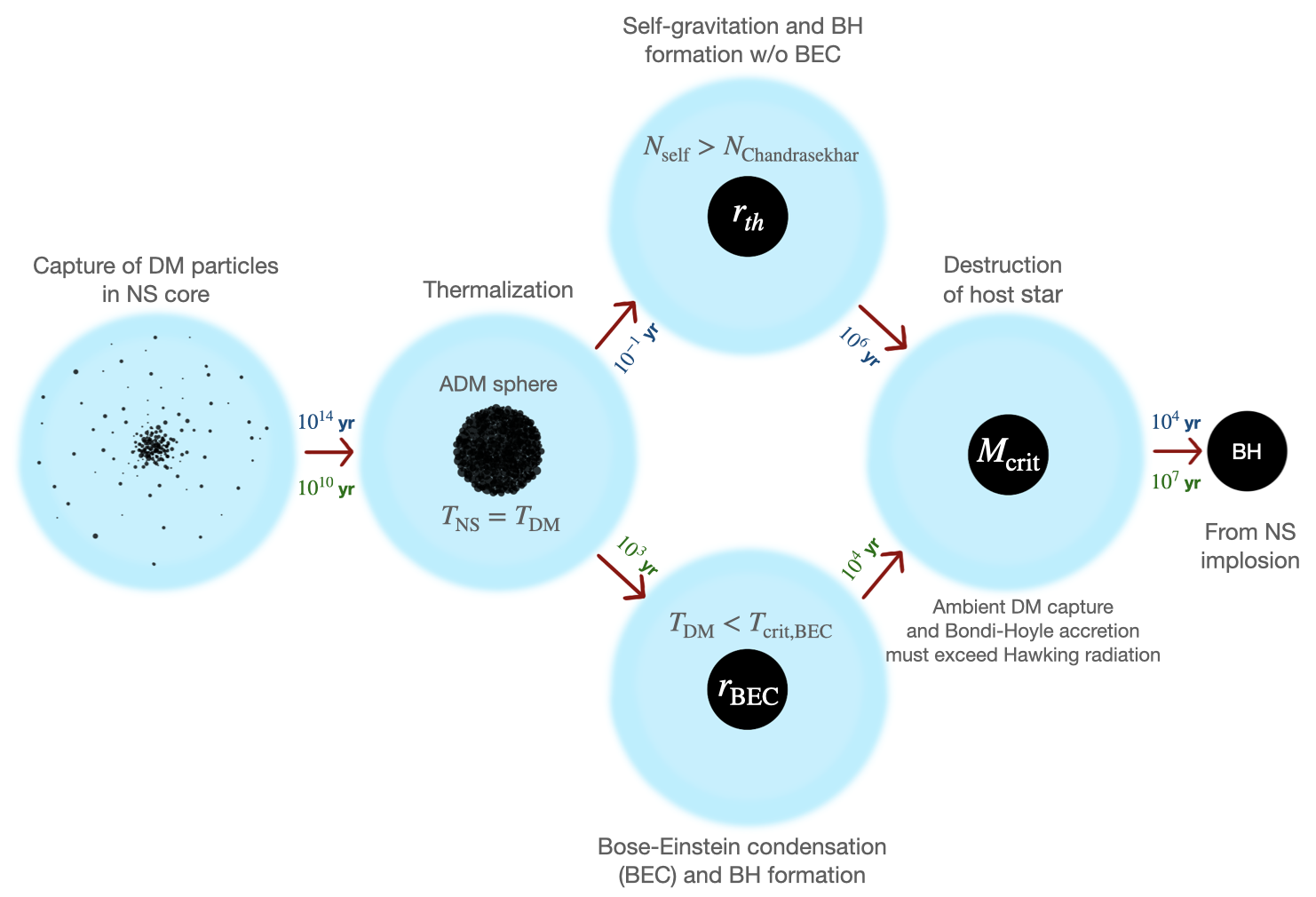}
    \caption{Two plausible scenarios for the formation of BHs by imploding NSs without significantly changing their mass. In the first scenario, DM particles accumulate, thermalize and form a self-gravitating object, which collapses to a BH if the number of DM particles exceeds the Chandrasekhar limit. In the second scenario, when a sufficiently large number of DM particles accumulates, a Bose-Einstein condensate (BEC) could form under favorable conditions and then collapse to a BH. Once a BH is assembled at the core, it can grow by accretion of NS matter, eventually leading to the implosion of the NS. The BH forms over a shorter timescale through BEC formation for DM particles of mass $m_\chi \leq 2 \times 10^4$ GeV. Therefore, we show the timescales with $m_\chi = 2 \times 10^4$~GeV for the channel where the BH forms without a BEC (blue), and with $m_\chi = 1$ GeV for the BEC channel (green), assuming a scattering cross-section $\sigma_\chi = 2\times10^{-45} \rm cm^{-2}$ and ambient dark matter density $\rho_\chi = 1\ {\rm GeV/cm^{3}}$.}
    \label{fig:implosion mechanism}
\end{figure*}

Observations of gravitational waves (GWs) by the Laser Interferometer Gravitational-Wave Observatory (LIGO) and Virgo over the past seven years~\cite{LIGOScientific:2018mvr,LIGOScientific:2020ibl,LIGOScientific:2021djp} have opened up a new avenue for exploring DM. On the one hand, detecting BHs of unusually large masses (compared to astrophysical BHs observed until then) could hint at their primordial origin~\cite{Bird:2016dcv}. This remains a possibility, although several astrophysical models can account for the broad range of BH masses detected by LIGO and Virgo (see, e.g.,~\cite{Wong:2020yig,Franciolini:2021tla}). The search for GWs from sub-solar mass BHs has so far been unsuccessful, severely constraining the fraction of total DM content in primordial BHs~\cite{LIGOScientific:2018glc, LIGOScientific:2019kan}. If the Universe has no primordial BHs with masses of ${\cal O}(10\,M_\odot)$, XG detectors can set upper limit on their abundance as a fraction of DM energy density as low as $f_{\rm PBH} \sim {\cal O}(10^{-5})$, about two orders of magnitude lower than current upper limits in this mass range; if instead $f_{\rm PBH} \gtrsim 10^{-4}$, future GW observations would exclude $f_{\rm PBH} = 0$ at the $95\%$ credible interval~\cite{Ng:2022agi}. The minimum testable abundance as a fraction of DM energy density depends on the primordial BH mass, and can be as low as $f_{\rm PBH} \sim {\cal O}(10^{-10})$~(see e.g. Fig.~5 of~\cite{DeLuca:2021hde}).

An alternative way to constrain the presence of DM would be to look for its gravitational drag on the orbits of BHs and neutron stars (NSs)~\cite{Eda:2013gg,Eda:2014kra,Hannuksela:2018izj,Hannuksela:2019vip,Kavanagh:2020cfn,Annulli:2020lyc,Traykova:2021dua,Coogan:2021uqv,Vicente:2022ivh,Speeney:2022ryg,Cole:2022ucw}. Additionally, the presence of an axionic cloud around BHs could extract the rotational energy from BHs, thereby affecting the spin distribution of BHs or producing continuous GWs from newly formed BHs~\cite{Arvanitaki:2010sy,Brito:2015oca}. Several authors have explored the prospect of making such observations~\cite{Arvanitaki:2014wva,Brito:2017wnc,Brito:2017zvb,Ng:2020ruv}. In fact, next-generation ground-based GW observatories, with the prospect of detecting several binary black hole (BBH) inspiral events each year with large signal-to-noise-ratios (SNRs), could observe dozens of post-merger axionic signals~\cite{Ghosh:2018gaw}, confirming or constraining bosons in the mass range $\sim [7 \times 10^{-14}, 2 \times 10^{-11}]$~eV~\cite{Yuan:2021ebu}.

In this paper, we explore the accumulation of bosonic DM in NS cores that could eventually form a stable, mini-BH, grow by Bondi-Hoyle accretion and eventually lead NSs to implode and form BHs, without significantly changing their mass. Two plausible scenarios are described in Fig.\,\ref{fig:implosion mechanism}. These mechanisms could be particularly efficient in regions of large DM densities, such as the central cores of large galaxies. It has been suggested that the lack of a sizeable population of pulsars in the core of the Milky Way, where the density of DM is expected to be particularly high, is because most of them have imploded to form BHs~\cite{Bramante:2014zca}. While this explanation might not be the root cause of why the Galactic center is deficient of pulsars, future GW observations could test if the implosion mechanism operates in NSs, as described below.

The timescale $t_c(\rho_\chi, m_\chi, \sigma_\chi)$ over which the accumulation of DM eventually makes NSs implode to form BHs depends on the DM density $\rho_\chi$ in the neighborhood of NSs, its mass $m_\chi$, and its interaction cross-section with hadrons $\sigma_\chi.$ NSs that live for a time longer than $t_c$ will get converted to BHs, and those that live for a shorter duration won't. Although isolated NSs can last forever, those in a merging binary would only live for a time $t_d$, called the \emph{delay time}, before they inspiral and merge to either form (rarely) supermassive NSs or (frequently) BHs. The delay time depends on the companion masses, and the periapsis and eccentricity of the binary at the time when it first forms. The delay time, therefore, is not the same for all binary neutron stars (BNSs). Instead, the NS binary population is characterized by a certain delay-time distribution $P(t_d)$. 

The probability distribution $P(t_d)$ is not well known, but it is often assumed to scale like $P(t_d) \propto 1/t_d$~\cite{Dominik:2012kk,Safarzadeh:2019znp,Safarzadeh:2019pis,McCarthy:2020jwq,Greggio:2020vyk}. The fraction of the population for which $t_d>t_c$ will be converted to BHs and the rest will remain as NSs. For the population of mergers detected with a sufficiently large SNR, GW observations can  determine the fraction of the BNS population that has been converted to BBHs. This fraction, if nonzero, can be used to infer the implosion timescale $t_c(\rho_\chi, m_\chi, \sigma_\chi)$ and hence constrain the parameter space of local DM density, DM mass, and interaction cross-section. If the population does not contain any BBHs, then it will be possible to set limits on the very same quantities. 

It is quite possible, although unlikely, that BHs in the mass range of NSs of 1--3 $M_\odot$ are produced by stellar evolution or, alternatively, they could be primordial in origin~\cite{Takhistov:2020vxs,Dasgupta:2020mqg}. The current consensus is that massive stars up to $\sim 23\,M_\odot$ leave behind NSs of masses in the range 1.2--2.0 $M_\odot$ at the end of their lives, while more massive stars are likely to leave behind a BH of mass greater than about $\sim 5\,M_\odot$~\cite{Bailyn:1997xt,2010ApJ...725.1918O,2011ApJ...741..103F,2012ApJ...757...91B,Fryer:2022lla}.
Although the primordial Universe could produce BHs in the mass range of NSs, they should also produce sub-solar mass BHs. A detection of sub-solar mass BHs could hint at the early-Universe origin of BHs with NS masses. Moreover, primordial BHs are expected to have small spin magnitudes if they do not increase their spin by coherent accretion~\cite{DeLuca:2020fpg}, while BHs formed from imploding NSs could have nonzero spins~(see, e.g.,~\cite{Gerosa:2018wbw,Belczynski:2017gds}). Consequently, it might be possible to discriminate between the two populations from their spin distributions~\cite{Franciolini:2021xbq}. Another proposal to distinguish primordial BHs from astrophysical compact objects is to use their mass distribution and the redshift evolution of the merger rates by using LVK and A+ detections of the stochastic GW background~\cite{Mukherjee:2021ags}, and possibly (in the future) Cosmic Explorer and Einstein Telescope observations of the stochastic background produced by sub-solar mass compact objects~\cite{Mukherjee:2021itf}.

The above argument assumes that either both NSs will implode, or neither does. If the time difference between the formation of the two NSs is large compared to the delay time $t_d$, then it is possible that only one of the NSs gets converted to a BH, but not its companion. However, this scenario is likely to be very rare.

\begin{figure}
\includegraphics[width=\columnwidth]{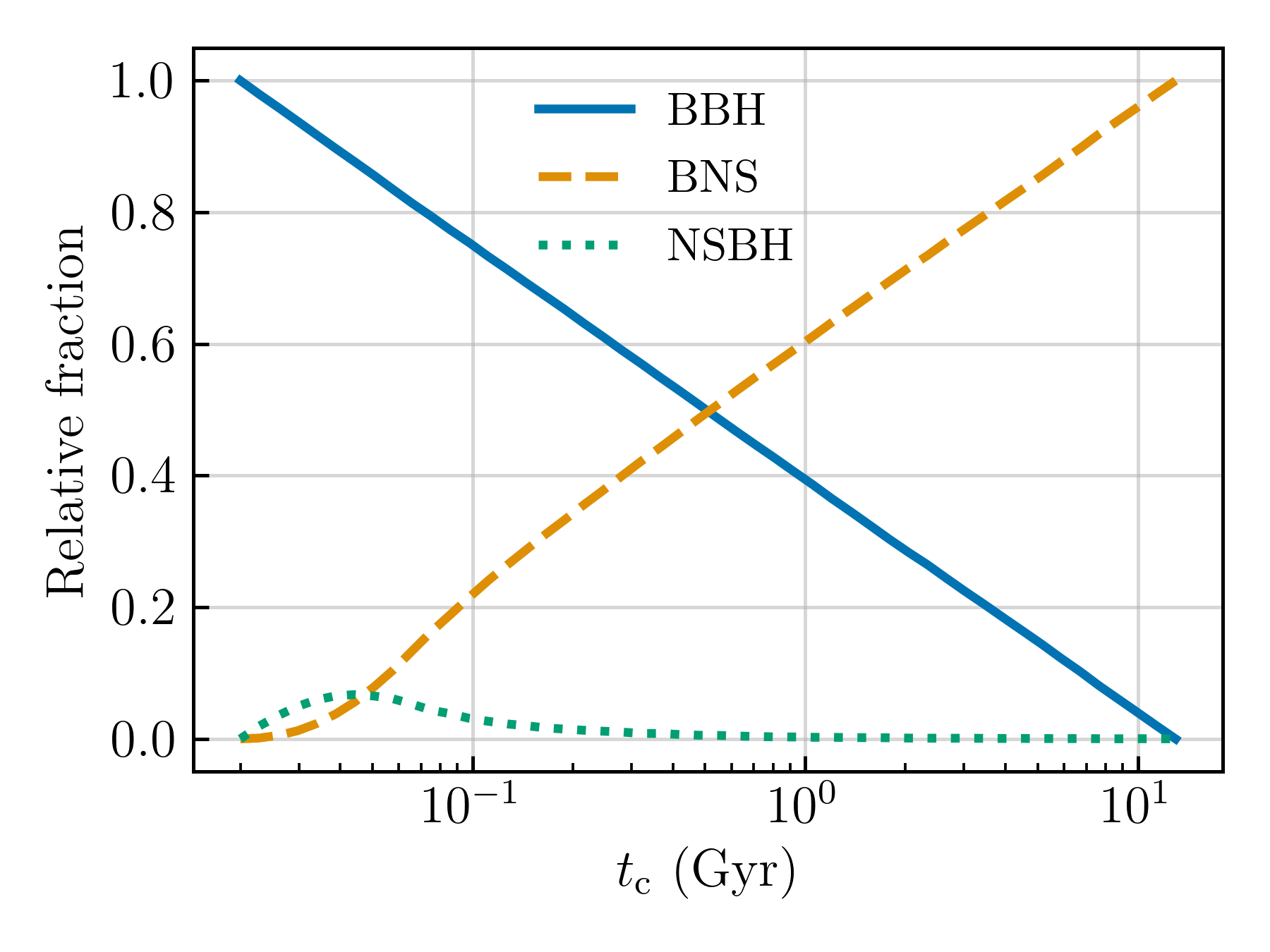}
\caption{The fraction of BBH and NSBH binaries formed from the implosion of one or both NSs, compared to the fraction of BNS systems as a function of the implosion time-scale $t_c$. As a function of $t_c$ the NSBH fraction remains negligibly small and can be ignored.}
\label{fig:fractions}
\end{figure}

The complex evolutionary process leading to NS formation is not completely understood, but stellar evolution models broadly suggest that NSs form from progenitors whose mass $M$ lies in the range $8\,M_\odot \lesssim M\lesssim 23\,M_\odot$~\cite{Fryer:2022lla}. 
The lifetime of such progenitor stars in the main-sequence, which varies as $10 (M/M_\odot)^{-2.5}\,\rm Gyr$~\cite{Hansen:1994}, would be in the range 4--55 Myr.  The heavier companion would evolve through the main sequence first to form a NS, followed by the lighter progenitor after a delay $\Delta t_{\rm MS}.$ Assuming that the progenitors are drawn from the Salpeter mass function, i.e., $P(m) \propto m^{-2.3}$~\cite{Chabrier:2004vw}, we find that $\Delta t_{\rm MS}$ has a median value of $\sim 14$ Myr, which is smaller than the smallest delay time $t_d^{\rm min}=20\,\rm Myr$ that we will be using in this work, and likely much smaller than the implosion timescale $t_{c},$ which could be as large as billions of years. A specific binary with delay time $t_d$ will be seen as a BBH if $t_c < t_d,$ as a BNS if $t_c > t_d + \Delta t_{\rm MS}$, or as a neutron star-black hole (NSBH) system otherwise.  In Fig.~\ref{fig:fractions} we plot the three fractions, and we see that the NSBH population constitutes at best about 6\% of the total, and only over a small range of values of $t_c.$ Consequently, we can safely assume that either both NSs in a binary would implode to form BHs, or neither would.

Observations of GWs could potentially discriminate the BBH population from that of BNSs. In the final moments before a BNS coalesces, each star experiences the tidal field of its companion, inducing a time-varying quadrupole deformation and associated emission of GWs. This is a high-order post-Newtonian effect -- technically a fifth post-Newtonian effect or a $(v/c)^{10}$ correction in the orbital phase evolution of the binary, where $v$ is the orbital speed and $c$ is the speed of light~\cite{Damour:1986ny,Mora:2003wt} -- which becomes important before the two NSs merge with each other, and will be absent in the case of BBHs~\cite{Damour:2009vw,Landry:2015zfa,Chia:2020yla,Poisson:2021yau}. Thus, by measuring the \emph{tidal polarizability,} often referred to as the \emph{effective tidal deformability}, it will be possible to ascertain the fraction of the two populations. It will not be possible to measure tidal deformability with sufficiently good accuracy for the entire observed population, but only for a fraction with sufficiently large SNR. In this paper we explore the sensitivity of future ground-based GW detectors to constrain the properties of a class of DM particles that can accumulate in NS cores and cause implosion, given sufficiently large time. 

The rest of the paper is organized as follows.  In Sec.~\ref{sec:tides}, we recall how tidal effects are encoded in GWs from BNSs. In Sec.~\ref{sec:fisher} we introduce the various GW detector networks and the waveform model used in this study, and compute the accuracy with which the tidal deformability of a BNS can be measured using the Fisher matrix formalism.  This will be followed by a computation in Sec.~\ref{sec:rates} of the relative rates of BBHs and BNSs, as a function of the unknown implosion timescale $t_c$ needed to form this novel population of BBHs. It turns out that the most important hyperparameter needed in the computation of relative rates is the implosion timescale $t_c.$ Thus, from the measured relative rates we can infer the implosion timescale, as shown in Sec.~\ref{sec:collapse time}.  In Sec.~\ref{sec:implosion}, we discuss the physics of implosion from DM accretion and the various timescales involved in the problem, from accumulation of DM particles to form a mini-BH, through self-gravitation with or without the formation of a Bose-Einstein condensate, its growth, and the final implosion of the NS (cf.\, Fig.~\ref{fig:implosion mechanism}).  In Sec.~\ref{sec:constraints} we derive the constraints that can be placed on DM particles if the proposed analysis does not find a single BBH event in the mass range of 1--3 $M_{\odot}$ (which would imply that the implosion timescale is larger than the Hubble time) and obtain the properties of bosonic DM particles assuming a collapse time of 1 Gyr.  In Sec.~\ref{sec:conclusions} we briefly summarize our findings, as well as our plans to apply this technique to the known population of LIGO-Virgo binary mergers for a number of implosion scenarios and different DM candidates.

\section{Tidal interaction to distinguish binary neutron stars from binary black holes}
\label{sec:tides}

In this Section we will discuss how to distinguish BNS mergers from BBH mergers using GW observations. An important difference in the GWs from the coalescence of BBH and BNS systems is that waves from BNS mergers have imprinted in them the tidal interaction between the two bodies, while BBH mergers will have no such signature~\cite{Damour:2009vw,Binnington:2009bb}. Additionally, while BBH mergers leave behind a BH remnant, BNS mergers could either promptly form a BH or leave behind a long-lived NS remnant with neutron-rich relativistic ejecta and a thermonuclear fireball. In this work, we will only consider the tidal interaction between the two bodies during the adiabatic inspiral regime. A merger accompanied by an electromagnetic afterglow essentially rules out a BBH merger.

\subsection{Tidal deformability}
A massive body produces a tidal field. The deformation induced by the tidal field on other bodies can be expressed as a multipole expansion, the quadrupole being the dominant multipole. Consider a spherically symmetric NS of radius $R$ in the tidal field ${\cal E}_{ij}$ of its companion NS. The quadrupole deformation $Q_{ij}$ induced in the star is related to the tidal field via the tidal deformability as 
\begin{equation}
    Q_{ij} = -\lambda {\cal E}_{ij},
\end{equation}
where $\lambda \equiv -\frac{2}{3G}k_2\,R^5,$ $R$ is the star's radius, and $k_2$ is the dimensionless tidal Love number. The tidal Love number, which measures a body's rigidity, depends on the equation of state (EOS) of the NS via its compactness $C\equiv Gm/(c^2R),$ where $m$ is the mass of the star~\cite{Chatziioannou:2020pqz}. For NS equations of state considered in this paper, we have $k_2\sim 0.1$~\cite{Hinderer:2009ca}.

In a binary system of stars orbiting each other the above quadrupole deformation is a function of time, which generates gravitational radiation, modifying the emitted signal at the fifth post-Newtonian order, inducing a $(v/c)^{10}$ correction to the dynamics of the system beyond the dominant quadrupole radiation reaction. In other words, the tidal interaction dissipates additional orbital energy into GWs, thus changing the orbital phase evolution of the waves at $(v/c)^{10}$ order beyond the quadrupole. This modification is significant in the final few cycles of the inspiral and coalescence of a binary, and can be detected if the signal is observed with a high SNR. 

The tidal deformability $\lambda$ has dimensions of $\rm kg~m^2~s^{2}$, but what appears in the post-Newtonian dynamics is the dimensionless tidal deformability, defined by
\begin{equation}
\Lambda \equiv \frac{c^{10}\lambda}{G^4m^5} = \frac{2}{3}k_2 {\cal C}^{-5}.
\end{equation}
As mentioned before, the Love number $k_2$ generally decreases with increasing compactness, thus the tidal deformability falls of steeper than $C^{-5}.$ For candidate equations of state of NSs $\Lambda$ decreases with the NS's mass and varies over the range $\sim [100, 4000]$  for NS masses in the range 1.1-1.5 $M_\odot$ considered in this study (see, e.g., Fig.\,1 of Ref.\,\cite{LIGOScientific:2019eut}), the smallest values corresponding to largest NS masses and softer equations of state, and largest values corresponding to smallest masses and stiffer equations of state. For BHs, $\Lambda=0$~\cite{Damour:2009vw,Landry:2015zfa,Chia:2020yla}: this is the key to distinguishing BBH mergers from NSBH and BNS mergers~\cite{Yang:2017gfb, Chen:2019aiw, Fasano:2020eum}.

\subsection{Tidal signature in neutron-star binary signal}
The GWs produced by BNSs are accurately described by post-Newtonian theory. We assume NSs have negligibly small spins and are on quasi-circular orbits. These are reasonable assumptions, as companions in Galactic double NS systems have negligible spins (based on Ref.\,\cite{Manchester:2004bp}, see also Fig.\,2.17 of Ref.\,\cite{Breton:2008yoi}) and gravitational radiation back reaction causes orbital eccentricity to decay more rapidly compared to the orbital separation \cite{Peters:1963ux}. In the Fourier domain, the strain amplitude $\tilde h(f)$ measured by an interferometric GW detector in response to an incident BNS signal on a quasi-circular orbit is given by:
\begin{equation}
    f\tilde h(f) = {\cal A}(f)\, e^{i\psi_{\rm PP}(f)+i\psi_{\rm Tidal}(f)},
\end{equation}
where $\psi_{\rm PP}$ and $\psi_{\rm Tidal}$ are contributions to the Fourier phase from the point-particle approximation and tidal effects, respectively, and a factor of $f$ is included to make the right-hand side dimensionless. The amplitude ${\cal A}(f)$ and the phase $\psi_{\rm PP}(f)$ are given by~\cite{Sathyaprakash:1991}:
\begin{eqnarray}
    {\cal A}(f) & = & \sqrt{\frac{5\nu}{6}} \frac{M}{D_{\rm eff}} \left ( \pi M f \right )^{-1/6},\\
    \label{eq:fourier phase}
    \psi_{\rm PP}(f) & = & \frac{5}{128\nu} \sum_{k=-5}^2 \left [ \alpha_k +\alpha_{kl}\log\frac{v}{v_0} \right ] (\pi M f)^{k/3}, \\
    D_{\rm eff} & = & \frac{4D_L}{\sqrt{F_+^2(1+\cos^2\iota)^2+ 4F^2_\times\cos\iota}}.
\end{eqnarray}
Here $M\equiv m_1+m_2$ is the binary's total mass, $m_1$ and $m_2$ are masses of the companion stars, $F_+(\theta,\phi,\psi)$ and $F_\times(\theta,\phi,\psi)$ are the detector antenna pattern functions [see, e.g.,~\cite{Sathyaprakash:2009xs}], $D_L$ is the luminosity distance to the source, $D_{\rm eff} > D_L$ is the effective distance, $(\theta, \phi)$ describe the position of the source in the sky, $\iota$ is the angle between the line sight to the binary and the orbital angular momentum, and $\psi$ is the polarization angle. The post-Newtonian coefficients $\alpha_k$ and $\alpha_{kl}$ depend on the symmetric mass ratio $\nu \equiv m_1m_2/M^2$, except for the log-terms in the post-Newtonian expansion and the tidal terms, which do depend on the total mass. The post-Newtonian expansion is carried out in powers of $v/c,$ where $v=(\pi M f)^{1/3}$, and $f$ is the GW frequency.  Relative to the dominant quadrupole term, the  tidal terms occur at the fifth post-Newtonian order and higher, i.e., a $(v/c)^{10}$ effect or, equivalently, $(\pi M f)^{10/3}$ term in the Fourier phase in Eq.~(\ref{eq:fourier phase}) relative to the dominant term~\cite{Flanagan:2007ix}.  The dominant tidal contribution to the phase and the first post-Newtonian corrections are (see e.g.~\cite{Favata:2013rwa}).
\begin{equation}
    \psi_{\rm Tidal}(f) = -\frac{39}{2}\tilde\Lambda v^{10} + \left ( \frac{6595}{364} \delta\tilde\Lambda - \frac{3115}{64}\tilde\Lambda \right ) v^{12},
\end{equation}
The dominant tidal term at the fifth post-Newtonian order depends on the mass ratio-weighted sum of the individual tidal deformabilities, defined as
\begin{subequations}
\label{eq:def_lams}
    \begin{equation}
    \tilde{\Lambda} = \frac{1}{26} \left[ (1+12q)\Lambda_{1} + \left (1+\frac{12}{q} \right) \Lambda_2 \right],
        \label{eq:lambdatilde}
    \end{equation}
while the correction at the sixth post-Newtonian order depends on $\tilde\Lambda$ as well as the difference in the tidal deformabilities, given by
    \begin{equation}
    \begin{aligned}
    \delta \tilde{\Lambda} &= \sqrt{1-4\nu}\left( 1-\frac{13272}{1319} \nu+\frac{8944}{1319} \nu^{2} \right) \frac{(\Lambda_2+\Lambda_1)}{2} \\ 
     &+\left( 1-\frac{15910}{1319} \nu+\frac{32850}{1319} \nu^{2}+\frac{3380}{1319} \nu^{3} \right) \frac{(\Lambda_2-\Lambda_1)}{2}.
    \end{aligned}
    \end{equation}
\end{subequations}
Note that for NSs of comparable masses, i.e., $\nu\simeq 1/4,$ we have $\Lambda_1 \simeq \Lambda_2,$ $\tilde\Lambda \simeq \Lambda_{1,2}$, and $\delta\tilde\Lambda \simeq 0.$ Thus, the second term is not only a sub-dominant post-Newtonian effect -- the expansion coefficient is itself small for most BNSs. Therefore, we neglect the sixth post-Newtonian correction in our computations.

While NSs have non-zero tidal deformability, the tidal deformability for BHs is zero. Tidal parameters can be inferred from GW signals, but current measurements have large uncertainties~\cite{De:2018uhw, LIGOScientific:2018cki}. Using GW observations, one can compute the value of $\Tilde{\Lambda}$ for the binary system and infer whether the system is a BNS or a BBH~\cite{Johnson-Mcdaniel:2018cdu, Chen:2020fzm, Fasano:2020eum}.

If the system is a BBH, it could have formed through the imploding DM channel. This allows us to find constraints on DM properties. On the other hand, if none of the systems are concluded to be BBH or NSBH binaries, we can still get limits for the DM properties. 

\section{Tidal deformability measurement with gravitational-wave detector networks}
\label{sec:fisher}
We now discuss the accuracy with which the effective tidal deformability $\tilde\Lambda$ can be measured using the GW signals emitted by coalescing BNSs. We will estimate the accuracy within the Fisher information matrix formalism~\cite{Finn:1992wt,Vallisneri:2007ev}, as implemented in {\sc gwbench}~\cite{Borhanian:2020ypi}. The two ingredients needed for the measurement of $\tilde\Lambda$ are (i) a GW detector network, which we introduce in Sec.~\ref{sec:network}, and (ii) the waveform model used in the Fisher matrix, which we discuss briefly in Sec.~\ref{sec:waveform}. Section~\ref{sec:lambda results} describes the accuracy with which $\tilde\Lambda$ can be measured with the detector networks considered in this study.

\subsection{Detector networks}
\label{sec:network}
Several authors have studied the capabilities of Advanced LIGO and Advanced Virgo in measuring the tidal deformability of NSs: for a review, see e.g.~\cite{Chatziioannou:2020pqz}. During the second observing run, LIGO and Virgo observed the first BNS inspiral event with a joint SNR of 33~\cite{LIGOScientific:2017vwq} which allowed the measurement of $\tilde\Lambda$ to within an accuracy of $\sigma_{\tilde\Lambda} = 630$ at 90\% credible interval \cite{LIGOScientific:2018hze}, with some authors ruling out the possibility that this was a BBH~\cite{De:2018uhw}, especially when combined with optical and infrared observations~\cite{Radice:2017lry}. The second BNS merger event, GW190425, was observed with a far lower SNR of 12.9, and it did not allow to place any meaningful bounds on the tidal deformability of NSs~\cite{LIGOScientific:2020aai}. The upcoming year-long fourth observing run of the LIGO, Virgo and KAGRA detectors is expected to detect a handful of BNS mergers but at the current sensitivity this network won't measure $\tilde\Lambda$ with an accuracy good enough to conclusively say that tidal effects are absent. We will therefore restrict ourselves to future upgrades of LIGO, Virgo, KAGRA and LIGO-Aundh,\footnote{LIGO-Aundh is the preferred name for the new LIGO-India observatory, coming up near the town of Aundh in central India.} as well as next generation (XG) observatories such as Cosmic Explorer~\cite{Reitze:2019iox} and the Einstein Telescope~\cite{Punturo:2010zza}.

More precisely, we consider three ground-based detector networks---A+, Voyager and XG---to determine the accuracy of measuring the effective tidal deformability, with particular interest in the XG network, since these observatories will have the required sensitivities to obtain an informative estimate of the tidal deformability parameter for BNSs.  
\begin{itemize}
    \item  The A+ network comprises five detectors: LIGO-Hanford, LIGO-Livingston, Virgo, KAGRA and LIGO-Aundh at A+ sensitivity~\cite{Barsotti:2018}.  
    \item The Voyager network consists of LIGO-Hanford, LIGO-Livingston and LIGO-Aundh at Voyager sensitivity~\cite{Adhikari:2019zpy}, with Virgo and KAGRA at A+ sensitivities.  
    \item Finally, the XG network includes the Einstein Telescope, one Cosmic Explorer in the US, and another Cosmic Explorer in Australia, as in~\cite{Borhanian:2020ypi}. 
\end{itemize}

\begin{table}[t!]
\centering
    \caption{\label{tab:Inj}Parameter space of BNSs used in this study. In addition to the detector-frame companion masses and the tidal deformability of NSs, there are four angles [two describing the orientation of the binary's orbit relative to the detector frame $(\iota,\,\psi)$ and two for the sky position of the source $(\alpha,\,\delta)$], the luminosity distance (or, equivalently, redshift $z$), a fiducial `arrival time' $t_C$ when the strain amplitude of the signal is largest, and the phase of the signal at that time $\phi_C$.}
\begin{tabular}{l c c}
\hline\hline
    Component mass, ${m_1}$ and ${m_2}$ & $\mbox{[1, 2]}\, M_\odot$ \\
    Tidal Parameters\footnote{These parameters don't directly enter the waveform, but only the effective tidal deformability $\tilde\Lambda.$}, $\Lambda_1$ and $\Lambda_2$ & 2.0 \\
    Effective tidal parameter, $\tilde\Lambda$ & $(6q^2+q+6)\Lambda_1/(13q)$\\
    Right ascension, $\alpha$ & [0,\, 2$\pi)$ \\
    Declination, $\delta$ & $[-\pi/2,\,\pi/2]$ \\
    Inclination, $\iota$ & [0,\, $\pi]$ \\
    Polarization, $\psi$ & [0,\, $2\pi]$ \\
    Redshift, ${z}$ & [0,\, 10] \\
    Fiducial time of arrival, $t_C$ & 0 \\
    Constant phase offset, $\phi_C$ & 0 \\
\hline\hline
\end{tabular}
\end{table}

\begin{figure*}
         \includegraphics[width=1.0\textwidth]{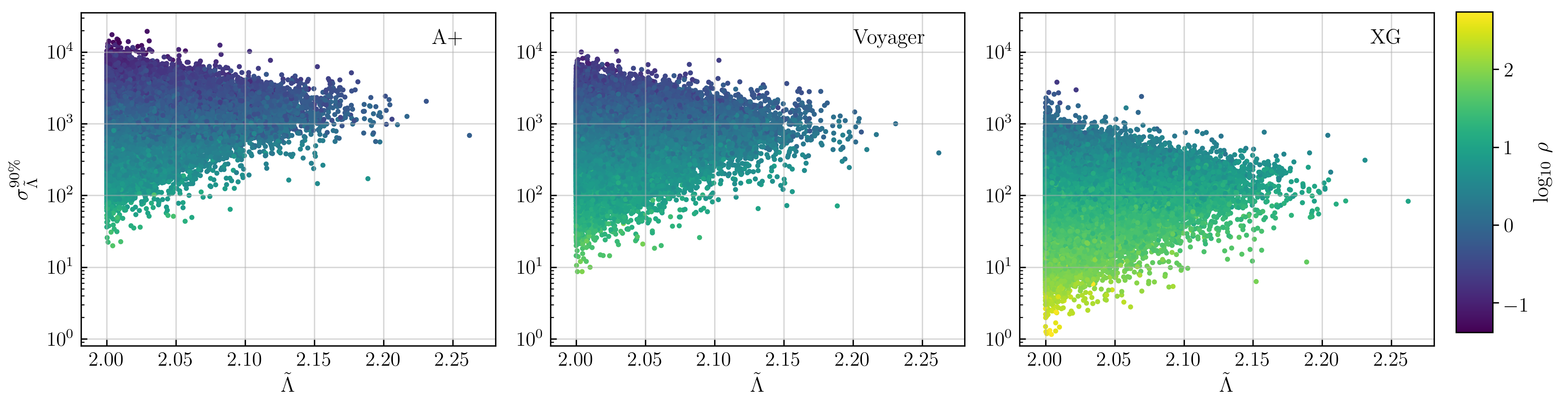}
        \caption{These plots show $\sigma_{\tilde{\Lambda}}^{90\%},$ the 90\% confidence interval in the accuracy of measurement of $\tilde\Lambda,$ for A+ (left panel), Voyager (middle panel) and XG (right panel), as a function of $\tilde{\Lambda}$ and the SNR (color bar). The measurement accuracy is an order of magnitude better for the XG network compared to the A+ network. In XG detectors, half of all events within $z=10$ have their tidal deformability constrained to within $\sigma^{90\%}_{\tilde{\Lambda}} \lesssim 100,$ an order of magnitude larger for than A+.}
        \label{fig:lamtilde inference and snr}
\end{figure*}

\subsection{Binary NS waveform model}
\label{sec:waveform}
As discussed in Sec.~\ref{sec:tides}, the signature of tidal deformation is imprinted in the GWs emitted by an inspiralling BNS system.  The simplest modification of the point-particle post-Newtonian model in the frequency domain -- cf.\, Eq.~(\ref{eq:fourier phase}) -- captures the essence of the tidal effects, but it is not in agreement with waveforms obtained from numerical simulations of BNS mergers. We adopt the \texttt{IMRPhenomD\_NRTidalv2} waveform model, in which the Fourier amplitude and phase are given algebraically in closed form~\cite{Dietrich:2017aum, Dietrich:2018uni, Dietrich:2019kaq}. This choice improves the accuracy in the calculation of derivatives of the waveform with respect to the various parameters, which are needed to compute the Fisher matrix.

The tidal terms in this model are valid at frequencies as low as $\sim 50$--100 Hz where Cosmic Explorer and Einstein Telescope have good sensitivity. The model does not incorporate the post-merger signal that could be present if the remnant is a long-lived NS, nor dynamical tides, which could also be important~\cite{Kokkotas:1995xe,Gualtieri:2001cm,Pons:2001xs,Berti:2002ry,Steinhoff:2016rfi,Andersson:2019dwg,Schmidt:2019wrl,Pratten:2021pro,Gamba:2022mgx,Williams:2022vct}, although sub-dominant compared to the static tides that are included in the model. 

The waveform model is described by ten parameters (see Table \ref{tab:Inj}): the companion masses $(m_1, m_2),$ two angles describing the position of the source in the sky $(\alpha, \delta)$, two angles describing the inclination of the binary's orbit and the wave polarization $(\iota, \psi),$ the luminosity distance of the source $D_L$  (or, equivalently, its redshift $z$), the effective tidal parameter $\tilde\Lambda$, the coalescence time $t_C$ and the coalescence phase $\phi_C$. 

In the Fisher matrix approximation, which involves derivatives with respect to the parameters of the waveform, the absolute error in the measurement of $\tilde\Lambda$ at leading post-Newtonian order is the same for all values of $\tilde\Lambda.$ This is because the effective tidal parameter appears linearly in the waveform model and we have neglected the sub-dominant tidal terms. Thus, the error $\sigma_{\tilde\Lambda}$ is determined by the correlation of $\tilde\Lambda$ with the other waveform parameters and the loudness of the signal and not any particular value of $\tilde \Lambda.$ To compute the Fisher matrix, without loss of generality we use a tidal deformability of $\Lambda_{\rm 1,2} = 2$, because  a nonzero tidal deformability value is required to use \texttt{IMRPhenomD\_NRTidalv2}.

The NS mass distribution is not known very well, but accurate measurement of NSs in radio binary pulsars seems to suggest that they are normally distributed with a standard deviation that's small compared to the mean (see~\cite{Kiziltan:2013oja, Farrow:2019xnc}). The two BNS mergers discovered by LIGO and Virgo, in particular GW190425~\cite{LIGOScientific:2020aai}, already indicate that NS masses in merging binaries could be different from those of galactic BNSs. There is currently no concrete distribution that we could use from GW measurements (see, however,~\cite{Landry:2021hvl}) and hence we draw the companion masses from a Gaussian distribution with a mean value of $1.3\ M_\odot$ and a standard deviation of $0.09\,M_\odot$~\cite{Farrow:2019xnc}, but with the constraint that $1\le m_1, m_2 \le 2 M_\odot.$ 

If $\Lambda_1=\Lambda_2=\Lambda,$ then Eq.\,(\ref{eq:lambdatilde}) leads to $$\tilde\Lambda = \frac{6q^2+q+6}{13q} \Lambda.$$ This implies that for $0.5\le q \le 1,$ which is the range of $q=m_2/m_1$ allowed by the Gaussian distribution and the hard upper- and lower-cutoff of NS masses, $2 \le \tilde\Lambda \lesssim 2.5.$ The range of $\tilde\Lambda$ is a proxy for the mass ratio in our sample, and the variation in the measurement error in $\tilde\Lambda$ is determined largely by its correlation with other parameters and by the SNR.

\subsection{Measurement accuracy of $\tilde\Lambda$}
\label{sec:lambda results}
Bayesian inference is the preferred method to estimate the error in the measurement of parameters, but the long Markov chains needed for the convergence of the posterior distribution are expensive and time consuming. For the exploratory work carried out in this paper it suffices to employ the faster Fisher matrix approach. In this approach, one first computes the Fisher information matrix $\Gamma_{mn}$ defined by
\begin{equation}
    \Gamma_{mn} \equiv \left < \partial_m \tilde h,\, \partial_n \tilde h \right >,
\end{equation}
where $\partial_m$ denotes the derivative of the waveform with respect to the parameter $\lambda_m$ of the waveform, $\tilde h(f)$ is the Fourier transform of the detector response $h(t)\equiv F_+h_+ +F_\times h_\times$, $(h_+,\,h_\times)$ are the two polarization strain amplitudes, and $\left <a,\,b\right >$ is the scalar product of waveforms $a$ and $b$, defined as
\begin{equation}
    \left <a,\,b\right > \equiv 4 \Re \int_{f_L}^{f_H} a(f)\,b^*(f) \frac{df}{S_h(f)}.
\end{equation}
Here $S_h(f)$ is the noise power spectral density of the detector in question, and $f_L$ and $f_H$ are suitably chosen lower and upper frequency cutoffs. We choose $f_L=20$ Hz for the A+ and Voyager networks, and $f_L=5$ Hz in the case of the XG network. The upper frequency cutoff is chosen to be the Nyquist frequency with a sampling rate of 4096 Hz. 

The information matrix of a detector network is just the sum of the individual information matrices: $\Gamma_{mn} \equiv \sum_A \Gamma^A_{mn}.$ The covariance matrix is the inverse of $\Gamma:$ $C_{mn} = (\Gamma^{-1})_{mn}.$ By definition the Fisher matrix is symmetric, and so is the covariance matrix.  Its diagonal elements $C_{mm}$ are the variances in the inference of parameters $\lambda_m$, and the off-diagonal elements $C_{mn}$ are the covariances in parameters $\lambda_m$ and $\lambda_n.$ 

We perform the analysis for a population of BBH sources with component masses $m_1, m_2 \in$ [\MinMass, \MaxMass], distributed in redshift as described in Sec.~\ref{sec:rates}, up to a maximum redshift of $z=10.$ Table~\ref{tab:Inj} lists the parameters of the binary population considered in this study. The errors on the effective tidal deformability $\sigma_{\Tilde{\Lambda}}$ for this population were computed using the {\sc gwbench} toolkit~\cite{Borhanian:2020ypi}, which performs Fisher analysis to provide measurement errors on GW parameters given a network of GW detectors. This provides the distribution of inferred values of $\Tilde{\Lambda}$ for the cosmic population of sources. We use $\sigma_{\Tilde{\Lambda}}^{90\%}$ as the criterion to differentiate between the population of BBHs and BNSs in this mass-range because $\tilde{\Lambda}=0$ for BHs. Therefore, the confidence with which we can classify a binary as a BBH or BNS is inversely proportional to the measurement error, given some $\Tilde{\Lambda}$. 

\begin{figure}[t]
    \includegraphics[width=0.99\columnwidth]{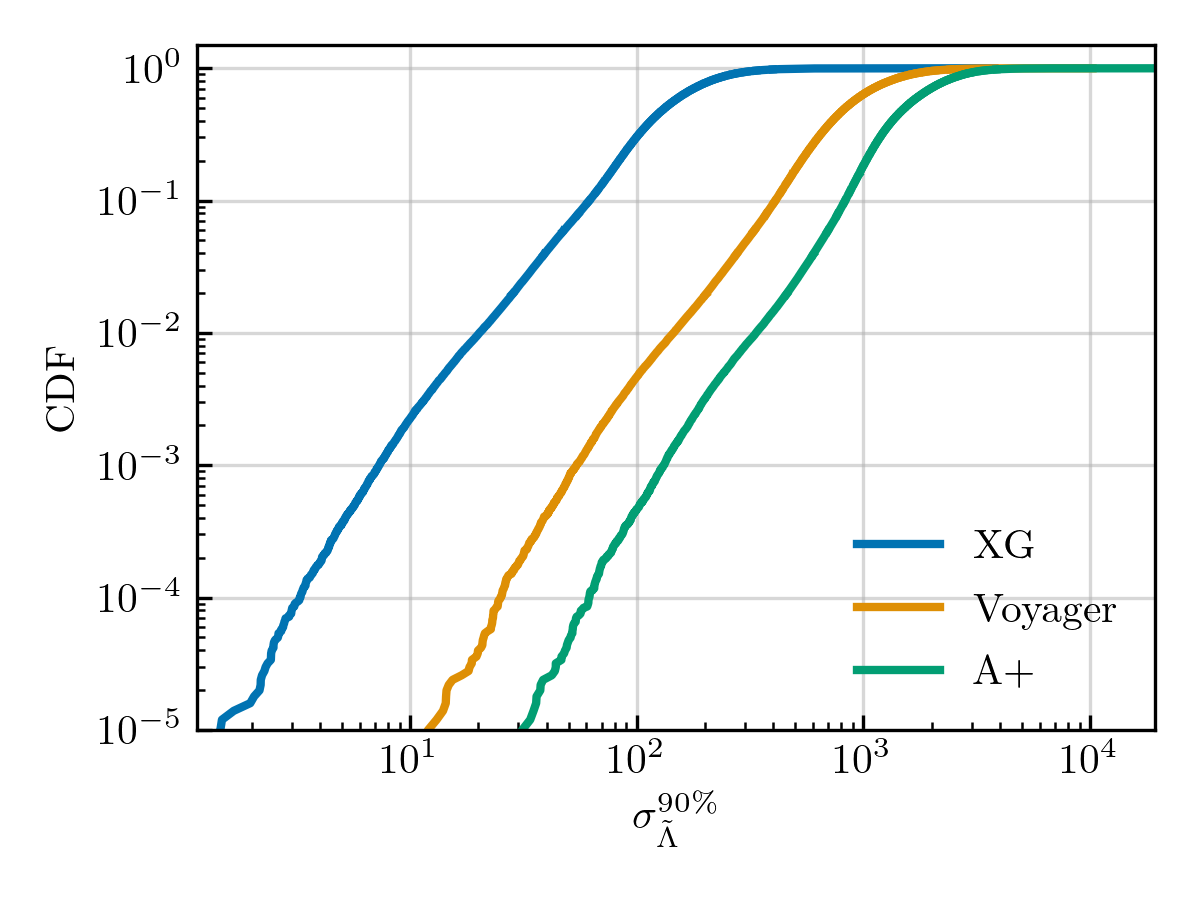}
    \caption{Cumulative distribution of the error in the measurement of effective tidal deformability in the three detector networks considered in this study. Notice that the fraction of events detected with an error $\sigma_{\tilde\Lambda}^{90\%} \simeq 100$ is 0.05\%, 0.5\% and 30\% in the A+, Voyager and XG networks, respectively.}
    \label{fig:lambda cdf}
\end{figure}

The 90\% confidence interval in the measurement of the effective tidal deformability for the full population, $\sigma_{\tilde\Lambda}^{90\%}$, is shown in Fig.~\ref{fig:lamtilde inference and snr} for the three networks considered in this study, with the color representing the SNR of the events.  In Fig.~\ref{fig:lambda cdf} we plot the cumulative distribution of $\sigma_{\tilde\Lambda}^{90\%}$ for the three networks. 
The A+ and Voyager networks can determine the tidal deformability to within $\sigma_{\tilde\Lambda}^{90\%} \simeq 100$ for 0.05\% and 0.5\% of the population. The XG network, on the other hand, can determine $\tilde\Lambda$ to the same accuracy for 30\% of the full population. This is a good enough accuracy to distinguish BNSs from BBHs if the preferred EOS is stiff, such as ALF2, producing larger NS radii and greater tidal deformabilities (e.g., $\tilde \Lambda \sim {\red 300}$ for a $1.4\,M_\odot$ NS, which is true for most NS masses considered in our study). Smaller errors $\sigma^{90\%}_{\tilde\Lambda} \lesssim {\red 20}$ would be required if NSs are described by a softer EOS, such as APR4, with tidal deformabilities $\tilde \Lambda\sim {\red 50}.$ A greater measurement accuracy (i.e., smaller values of $\sigma^{90\%}_{\tilde\Lambda}$) requires louder signals, which means fewer systems can be classified as belonging to one of the two classes.
For example, only the XG network can measure $\tilde\Lambda$ to better than $\sigma_{\tilde\Lambda}^{90\%}<20.$ Thus, the measurement accuracy of effective tidal deformability directly impacts how well we can determine the collapse time discussed in Sec.~\ref{sec:collapse time}.

\section{Merger rates}
\label{sec:rates}
In this section we discuss how to deduce constraints on DM mass and interaction cross-section based on the observed merger rate of BNS and BBH systems in the NS mass range. The constraints follow by comparing the time-scale for conversion of NSs to BHs by DM accumulation, $t_c,$ to the time-scale for coalescence of NSs by gravitational radiation backreaction, $t_d$.
The timescales in the problem suggest that either both or neither of the NSs in a binary will be converted to BHs before they inspiral and merge, so we will not consider NSBH binaries, but we will derive an equation that relates an upper limit on the merger rate of BBHs to the properties of DM particles.

\begin{figure*}
    \centering
    \includegraphics[width=\columnwidth]{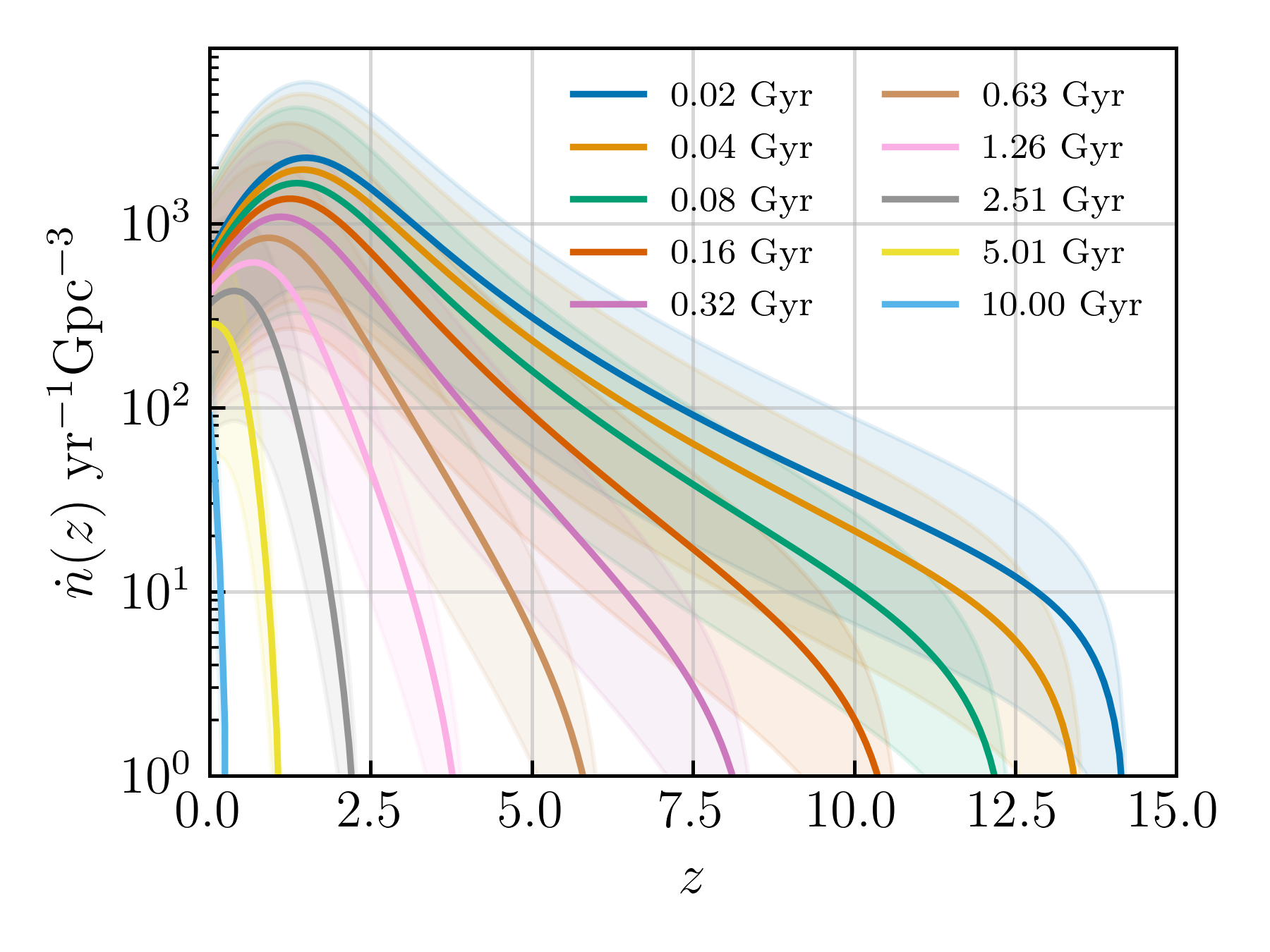}
    \includegraphics[width=\columnwidth]{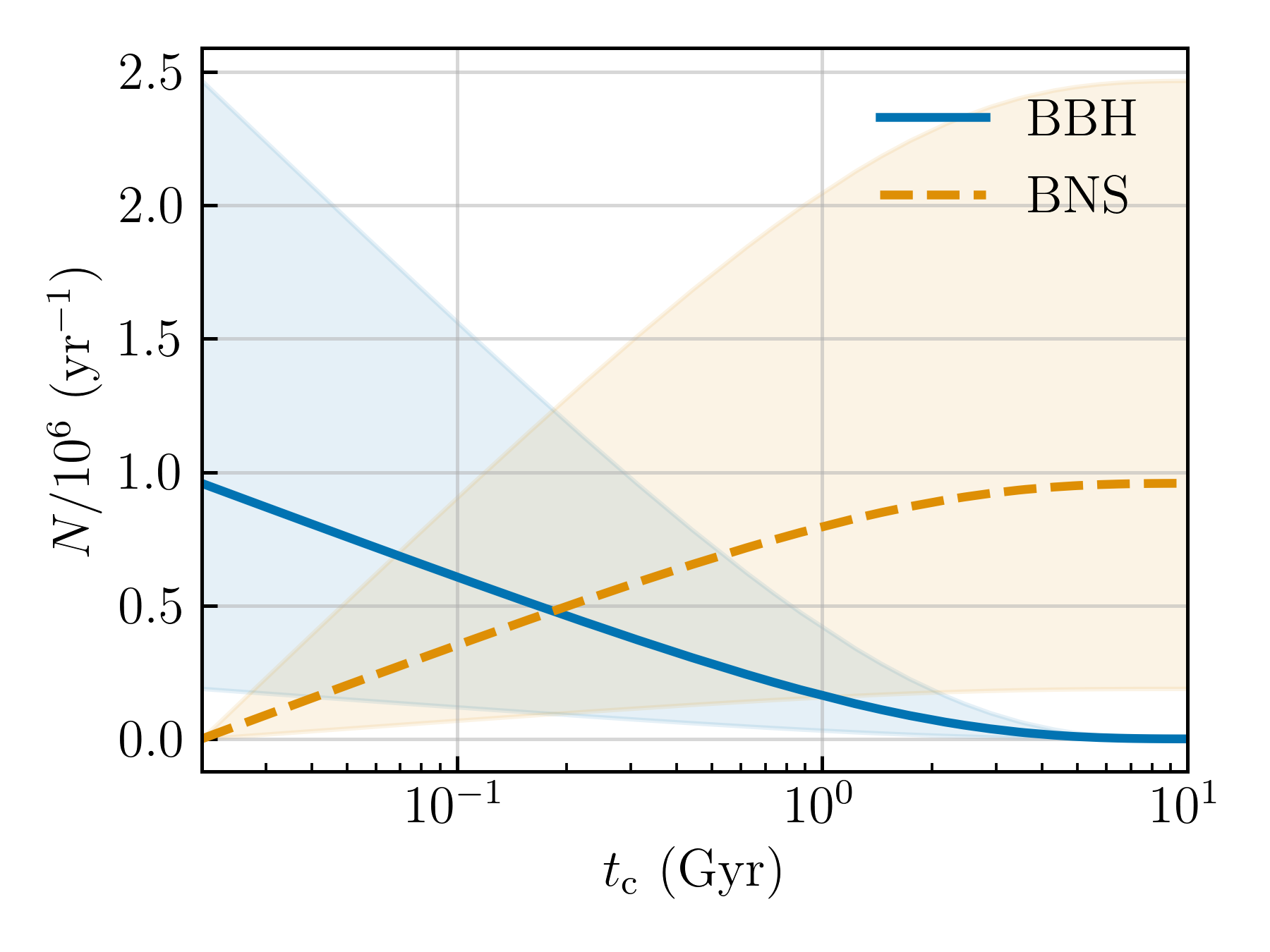}
    \caption{Left: Merger rate density of BBHs formed by implosion as a function of redshift for several values of the collapse time $t_{c}$ from 0.02 Gyr to 10 Gyr, as given in the legend (in Gyr).
    Right: The number of BNS and BBH mergers per year $N = \dot N T,$ $T=1\,\rm yr,$ in Earth's frame as a function of the collapse time $t_{c}.$ Not all of these would be observable by a detector network, but only those that above a certain SNR threshold. The shaded regions in both panels correspond to the 90\% credible interval for the local BNS merger rate for the multi-source model in~\cite{LIGOScientific:2021psn}.}
    \label{fig:local rate}
\end{figure*}

\subsection{Collapse timescales and merger time delays}
\label{sec:collapse time scales and merger time deilays}
As we shall discuss in Sec.~\ref{sec:implosion}, the time-scale $t_c$ for NSs to implode to form BHs largely depends on the properties of the DM particles: (i) the interaction cross section $\sigma_\chi,$ (ii) the DM density at the site of BNSs and their dispersion velocity (which are both determined by the location of the binary within a galaxy, being larger at the galactic core and smaller in the halo), and (iii) the mass of the DM particles $m_\chi.$ On the other hand, the time-scale $t_d$ for NSs to coalesce depends on (i) the eccentricity and semi-major axis when the BNS first forms, and (ii) the masses of the two NSs~\cite{Peters:1963ux}.

The two NSs in a binary do not form from their stellar progenitors at the same time. The delay in the formation of the second NS with respect to the first could be substantial if the masses of the parent stars are very different.  
However, the timescale arguments in Sec.~\ref{sec:intro} imply that binaries with component masses $1\,M_\odot\le m \le 2\,M_\odot$ will either be BNS or BBH binaries, and not mixed (NSBH) binaries, so we will ignore mixed binaries from now on.

\subsection{Binary neutron star merger rate}
LIGO and Virgo have so far observed two BNS mergers: GW170817 at a distance of $40^{+8}_{-14}$ Mpc~\cite{LIGOScientific:2017vwq} and GW190425 at a distance of $159^{+69}_{-72}$ Mpc~\cite{LIGOScientific:2020aai}, both at 90\% credible interval (CI). Since they are both at very low redshift, the merger rate determined from them is essentially the local (i.e., $z=0$) rate. The local rate for BNSs inferred from the third Gravitational Wave Transient Catalog-3 GWTC-3~\cite{LIGOScientific:2021djp} is $R_0 = 660_{-530}^{+1040} {\rm Gpc^{-3}\,yr^{-1}}$ at 90\% CI~\cite{LIGOScientific:2021psn}, under the assumption of a multi-source model including BNS, NSBH, and BBH sub-populations.  

The merger rate evolves with redshift because (i) the star formation rate varies as a function of redshift; (ii) BNSs that form at a certain redshift don't merge immediately but only after a delay time $t_d$, and hence at a different redshift; and (iii) the metallicity evolves with redshift, affecting the mass function and formation rate of compact binaries. In this study, we will ignore the effect of metallicity, as it plays a greater role in the case of BBHs and is less likely to affect the merger rate of BNSs~\cite{Dominik:2012kk,Dominik:2013tma,Dominik:2014yma,Santoliquido:2020axb}. We will assume the star formation rate $\psi(z)$ (SFR) given by Ref.~\cite{Madau:2014bja}: 

\begin{equation}
    \psi(z | \alpha_{\rm{F}}, \beta_{\rm{F}}, C_{\rm{F}}) 
    \propto
    \frac{(1+z)^{\alpha_{\rm{F}}}}{1+\left(\frac{1+z}{C_{\rm{F}}}\right)^{\beta_{\rm{F}}}}\,,
\end{equation}
with $(\alpha_{\rm{F}}, \beta_{\rm{F}}, C_{\rm{F}}) = (2.7,5.6,2.9)$. We assume that the merger rate is the same as the SFR except that binaries that form at redshift $z_f$ merge at redshift $z$ after a delay $t_d,$ with a corresponding redshift interval $\Delta z=z_f-z.$ Given a redshift $z$ at which merger rate is required and the time delay $t_d,$ the redshift $z_f$ at which the binary forms can be found by solving
\begin{equation}
    t_d = \frac{1}{H_0} \int_z^{z_f}\frac{dz'}{(1+z')E(z')},\quad E(z) = \Omega_\Lambda + \Omega_M(1+z)^3,
    \label{eq:zf}
\end{equation}
where $\Omega_M$ and $\Omega_\Lambda$ are the DM and dark energy densities, respectively, and we have assumed a flat Universe in which dark energy is interpreted as a cosmological constant~\cite{Sahni:1999gb}.  Now, the merger rate density $\dot n(z)$ as a function of redshift can be computed by integrating the SFR over all delay times, the delay-time probability function $P(t_d)$ serving as a weighting factor:
\begin{gather}
    \dot{n}(z) = A \int_{t_d^{\rm min}}^{t_d^{\rm max}} \psi(z_f(z,t_d)) \mathcal{P}(t_d)\, dt_d, 
    \label{eq:ndot}
\end{gather}
where $z_f$ is found using Eq.\,(\ref{eq:zf}). The delay-time distribution is not very well known, and we take it to be the Jeffrey's prior $\mathcal{P}(t_d)\propto t_d^{-1}$ extending over the range $[t_d^{\rm min},\, t_d^{\rm max}],$ with the normalization $$\int_{t_d^{\rm min}}^{t_d^{\rm max}} P(t_d)\,dt_d=1.$$  The constant $A$ in Eq.~(\ref{eq:ndot}) is chosen such that $\dot n(0)=R_0:$ 
\begin{equation}
    A = \frac{R_0}{\int_{t_d^{\rm min}}^{t_d^{\rm max}} \psi(z_f(0,t_d))P(t_d)\,dt_d}.
\end{equation}

\begin{figure*}[t]
     \centering
        \includegraphics[width=\textwidth]{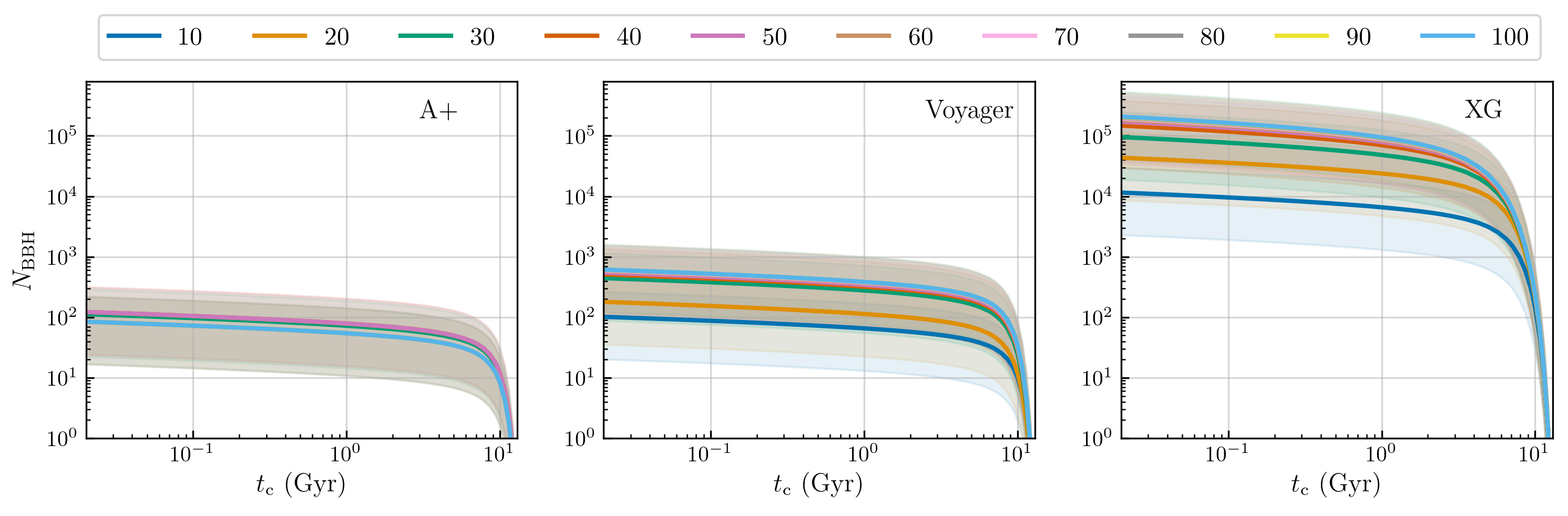}
        \caption{The number of BBH mergers as a function of collapse time $t_{c}$ for several values of the threshold $\sigma_{\Tilde\Lambda_T}$ on the estimated error in $\tilde{\Lambda},$ for 5 years of observing time. }
        \label{fig:NBBH}
\end{figure*}

If NSs in a binary implode to form BHs within a collapse time $t_{c}$ that lies between the minimum and maximum delay times, i.e., $t_d^{\rm{min}} < t_{c} < t_d^{\rm{max}}$, then the total merger rate density in Eq.\,(\ref{eq:ndot}) is the sum of the BNS and BBH merger rates, i.e., $\dot{n}_{\rm total} = \dot{n}_{\rm BNS} + \dot{n}_{\rm BBH}$, where 
\begin{eqnarray}
    \dot{n}(z)_{\rm BNS} & = & A \int_{t_d^{\rm min}}^{t_{c}} \psi(z_f(z, t_d)) \mathcal{P}(t_d)\, dt_d, \\
    \label{eq:ndot bns}
    \dot{n}(z)_{\rm BBH} & = & A \int_{t_{c}}^{t_d^{\rm max}} \psi(z_f(z, t_d)) \mathcal{P}(t_d)\, dt_d.
    \label{eq:ndot bbh}
\end{eqnarray}
The left panel of Fig.~\ref{fig:local rate} shows the merger rate of BBHs formed by implosion as a function of redshift for different choices of the collapse time $t_{c}$ (in Gyr). The shaded region for each choice of $t_{c}$ represents the uncertainty in the local merger rate of BNS found using the latest GW catalog, which is in the range  $130 \le R_0 \le 1700\,{\rm Gpc^{-3}\,yr^{-1}}$~\cite{LIGOScientific:2021psn}. If the collapse time is less than the smallest delay time, then all BNSs are converted to BBHs, and the entire BNS population will be observed as BBHs. If, on the other hand, the collapse time is larger than the largest delay time, no BNS is converted to BBH. The observed population will be a mixture of BNS and BBH for values of the collapse time in between these extremes.

The merger rate $d\dot N$ in a cosmological volume $dV_c$ (redshift range $dz$), as measured by an observer at $z=0,$ is $\dot N = \dot n(z)\,dV_c/(1+z),$  where a factor $(1+z)$ accounts for the time dilation between observers at redshift $z$ and redshift $0.$ Thus, the merger rate within some redshift $z$ as measured by an observer at $z=0$ is given by 
\begin{equation}
    \dot{N} = \int_0^z \frac{\dot{n}(z')}{1+z'}\frac{dV_c}{dz'}\,dz'.
    \label{eq:Ndot}
\end{equation}
In our simulations we distribute sources as a function of redshift using the above equation.  The right panel of Fig.~\ref{fig:local rate} shows the number of BNS and BBH mergers in the Universe per year $N=\dot NT,$ with $T=1\,\rm yr,$ in the Earth's frame, as a function of the collapse time $t_{c}.$ If $t_{c}<t_d^{\rm min}$ every BNS will be converted to a BBH, while $t_{c}>t_d^{\rm max}$ would imply none will be converted. At intermediate values of $t_{c}$ we expect to observe, among all mergers, some that are BBHs.  

A detector network only observes those mergers that stand above the detector noise at an acceptably low false alarm rate (say, one false event per year). Thus, the merger rate \emph{observable} by a detector network, $\dot N_{\rm obs}$, is
\begin{equation}
    \dot{N}_{\rm obs} = \int_0^z 
    \frac{\dot{n}(z')}{1+z'}\frac{dV_c}{dz'}\epsilon(z')dz'.
    \label{eq:Ndet}
\end{equation}
Here $\epsilon(z)$ is the detection efficiency of a network as a function of redshift, to be discussed below.  It measures the fraction of events detected by a given detector network subject to one or more conditions. At the outset, we require an event's SNR to be larger than a preset threshold $\rho_T$ to make a high-confidence detection. For the current network of LIGO/Virgo/KAGRA detectors, a SNR threshold of $\rho_T=10$ assures that the false alarm rate is no more than a few per year. 

In this work we not only wish to make high-confidence detections, but additionally select only those events for which the effective tidal deformability can be measured with a good accuracy---good enough to distinguish BBH mergers (for which $\Tilde\Lambda=0$) from BNS mergers (for which $\Tilde\Lambda>\Tilde\Lambda^{\rm NS}_{\rm min}\gg 0$). To this end, we require the 90\% credible interval $\sigma_{\Tilde{\Lambda}}^{90\%}$ in the measurement of $\Tilde\Lambda$ to be less than a preset threshold $\sigma_{\Tilde\Lambda_T}$, i.e.,  $\sigma_{\Tilde{\Lambda}}^{90\%} < \sigma_{\Tilde \Lambda_T}$, and choose the threshold so that it is (significantly) smaller than the smallest value of the effective tidal deformability of BNSs:   
$$\sigma_{\Tilde\Lambda_T} \ll \Tilde\Lambda^{\rm NS}_{\rm min} \equiv \min_{m_1,m_2} \Tilde{\Lambda}(m_1, m_2),$$  
where $m_1$ and $m_2$ are the companion masses. This  condition guarantees that a merger is correctly classified as a BNS or a BBH merger with high confidence. Typically, NSs with the largest masses and softest equations of state have the smallest tidal deformability. For equations of state that are still viable, the smallest value of  $\Tilde\Lambda$ varies over the range $\Tilde\Lambda \in [10, 200]$~\cite{LIGOScientific:2019eut}. We present our results for several values of the threshold  $\sigma_{\Tilde\Lambda_T}$ restricting the value to be less than 100. 

With the conditions on SNR and $\Tilde\Lambda$ imposed, the efficiency of a detector network is given by:
\begin{equation}
\epsilon(z) = \frac{1}{N} \sum^{N}_{i=1} \Pi\left(\frac{\rho}{\rho_T}-1 | z\right) \Pi\left(\frac{\sigma_{\Tilde\Lambda_T}}{\sigma_{\Tilde{\Lambda}}} - 1 | z\right),
\label{eq:epsilon}
\end{equation}
where the sum is over the full population of BNSs and $\Pi(x)$ is the step function: $\Pi(x)=0$ if $x\le 0$, and $\Pi(x)=1$ if $x>0$.  

\begin{figure*}
     \centering
         \centering
         \includegraphics[width=0.49\textwidth]{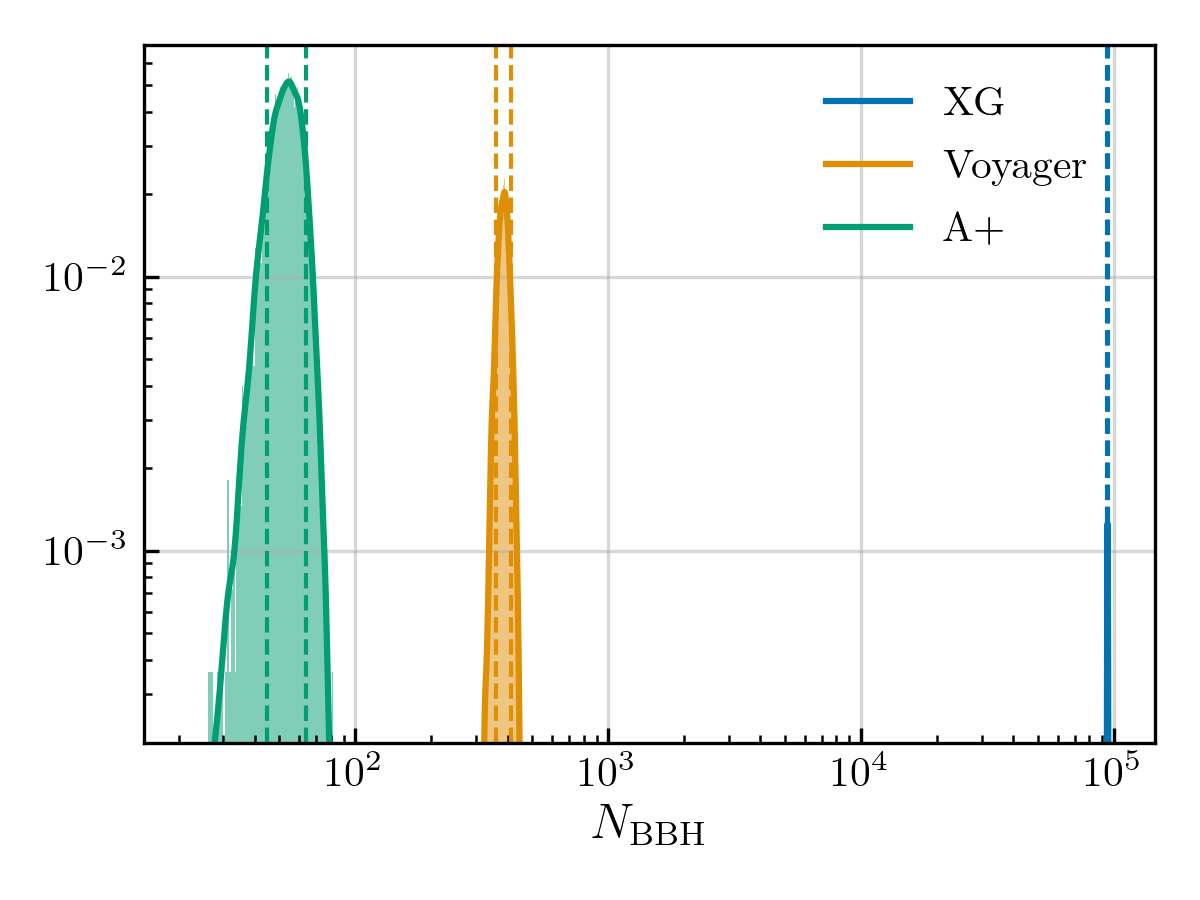}
         \includegraphics[width=0.49\textwidth]{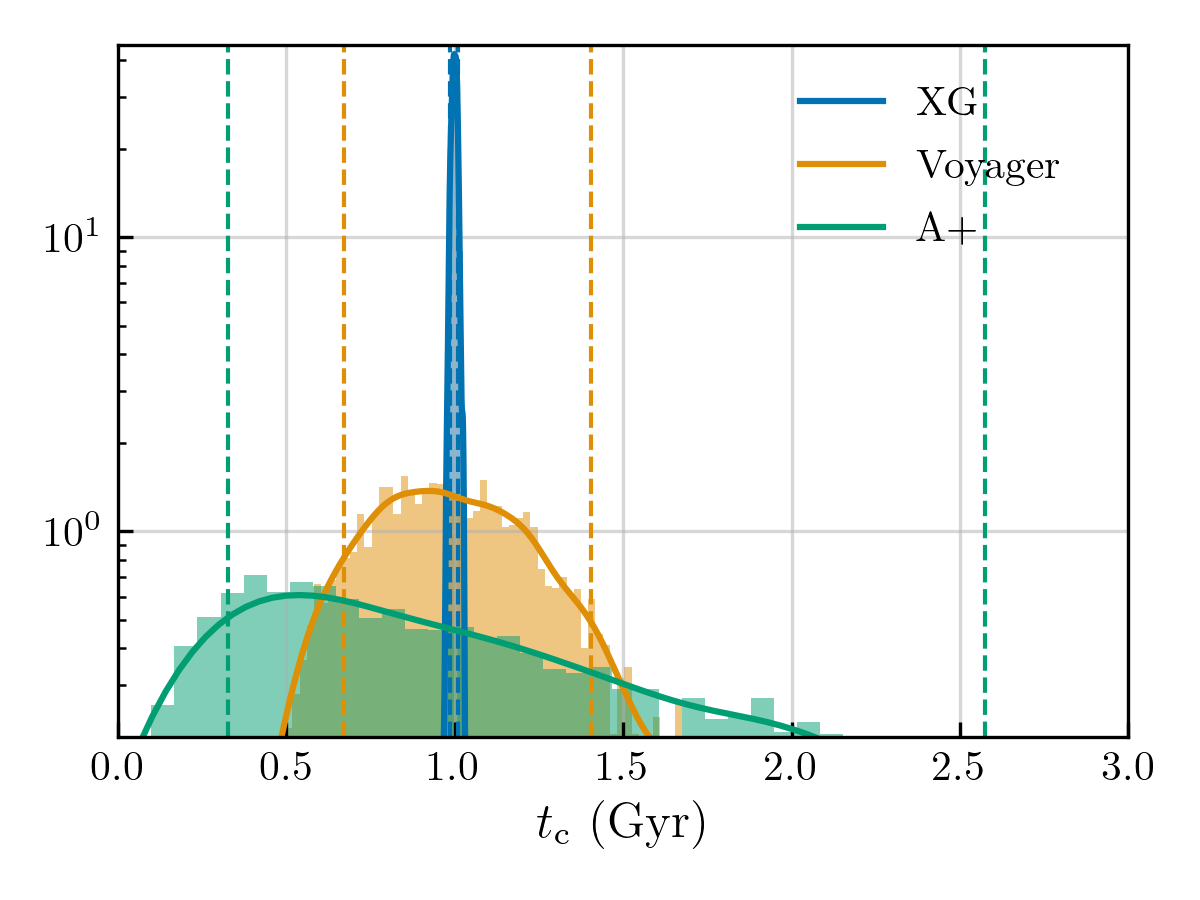}
         \label{fig:five over x}
        \caption{Left: Gaussian distribution for the observed number of BBHs for $\sigma_{{\tilde{\Lambda}_T}}^{90\%} = 100$ and $t_{c}=1~{\rm Gyr}$. 
        Right: Inferred distribution of $t_{c}$ from $N_{\rm BBH}$ for the three detector networks under consideration and an observing period of 5 years. The dashed lines show the 90\% credible intervals for the distributions of $t_{c}$ from $N_{\rm BBH}$.}
        \label{fig:NBBH and tcollapse}
\end{figure*}

The number of BBH mergers observed up to some redshift, $z$, in an observing period $T,$ is given by 
\begin{equation}
    N_{\rm BBH} = \dot N_{\rm BBH} T = T \int_0^z \frac{\dot{n}(z')_{\rm BBH}}{1+z'}\,  \frac{dV_c}{dz'}\, \epsilon(z')dz',
    \label{eq:NBBH}
\end{equation}
and similarly for BNSs. Therefore, the number of BBHs we expect to observe in this mass spectrum not only depends on the sensitivity of GW detectors and searches, but also on how well we can measure the tidal effects from these observations. A single detection with some assumed measurement efficiency for $\tilde{\Lambda}$ provides a limit on $t_{c}$ which can be used to infer DM properties using Eqs.~\eqref{eq:tc accretion} and \eqref{eq:tc BEC}.

Figure~\ref{fig:NBBH} shows the number of BBH mergers \emph{observable} by a detector network over a five-year duration, $N_{\rm BBH} = \dot N_{\rm BBH} T$ with $T=5\,\rm yr,$ as a function of the collapse time $t_{c}$ for different choices of the threshold $\sigma_{\Tilde\Lambda_T}$ (from 10 to 100) and for the A+, Voyager and XG detector networks described in Sec~\ref{sec:fisher}. The imminent upgrade of LIGO and Virgo detectors could identify tens to hundreds of of BBHs if the collapse time is in the range of 100 Myr to 1 Gyr over an observational period of 5 years but this network is more likely to acquire about a year's worth of data. The Voyager network increases these numbers by an order of magnitude while the XG network of Einstein Telescope and Cosmic Explorer will observe several orders of magnitude more BH binaries compared to the other networks. As we shall see later, the larger fraction of systems that can be clearly identified as BBH helps in placing a tighter constraints on DM mass and interaction cross section.

\section{Inference of collapse time}
\label{sec:collapse time}
In this section we elucidate how to infer the average time it takes for a NS to collapse to a BH due to accretion of DM. Let us first note a caveat in our argument: what GWs \emph{can} infer is the relative abundance of BBHs versus BNSs. If BHs with companion masses in the 1--2 $M_\odot$ range form by unknown astrophysical processes or in the primordial Universe they will be part of the BH population in this mass range, and will be  indistinguishable from the population that formed from the implosion of NSs due to accumulation of DM. First, we assume that stellar evolution cannot produce BHs in this mass range. Second, if primordial BHs exist then there is no fundamental reason they should only appear in this mass range. In particular, we expect sub-solar mass BBHs to exist as well~\cite{Byrnes:2018clq, LIGOScientific:2021job}. In the absence of such a population we can be fairly confident (although not certain) that stellar mass BHs in the 1--2 $M_\odot$ range are not of primordial origin. At present, there is no preference for any of these scenarios, so we will proceed with the assumption that any detections of BBHs with BNS masses formed by the implosion of NSs.

Under this assumption, it is straightforward to deduce the collapse time from the observed population of BBHs and BNSs. For illustration, we assume that the true collapse time is $t_c=1\,\rm Gyr.$ Given $t_c,$ Eq.~(\ref{eq:ndot bbh}) gives the local merger rate density, which could then be used in Eq.~(\ref{eq:NBBH}) to compute the number of BBHs expected to be detected over an observational period $T$ in a detector network whose detection efficiency is $\epsilon(z)$, as defined in Eq.~(\ref{eq:epsilon}).  Since the rate is Poisson distributed, the number of detections $N_{\rm BBH}$ would be uncertain by $\sqrt{N_{\rm BBH}}$, and the relative error scales like $1/\sqrt{N_{\rm BBH}}.$  

The left panel of Fig.\,\ref{fig:NBBH and tcollapse} shows the expected number of BBHs in different detector networks, together with the rate uncertainty in our simulation.  The relative uncertainty will, obviously, be larger for less sensitive detectors, and this impacts how well the collapse time can be deduced. The right panel of Fig.~\ref{fig:NBBH and tcollapse} shows the collapse time deduced from the rate posterior plotted on the left. In reality, we would determine the number of BBH mergers given the collapse time and network efficiency [i.e., $N_{\rm BBH}(t_c, \epsilon(z))]$ and interpolate this function to determine $t_c$ from $N_{\rm BBH}$ for a given network efficiency $\epsilon(z).$ It is clear that XG detectors will be able to constrain the collapse time far better than the A+ network.

\section{Implosion of neutron stars by accumulation of dark matter}
\label{sec:implosion}
The inferred collapse time of NSs into BHs from GW observations is model-agnostic. These limits can constrain the particle properties of DM in scenarios where DM particles get captured in NS cores. In this section, we illustrate how we can constrain the asymmetric DM scenario under certain simplified assumptions using previously derived limits on the collapse time.

In the asymmetric DM scenario, DM particles do not self-annihilate due to the assumed asymmetry between the number density of particles and anti-particles. Therefore, the capture and accumulation of DM particles in the core of NSs could lead to the formation of a BH at the NS core, leading to the potential implosion of the host star. Here we examine this scenario in the case of non-interacting bosonic DM, because the bosonic Chandrasekhar limit is much greater than that for fermions, implying that bosonic DM undergoes gravitational collapse sooner than fermionic DM. Indeed, the number of fermionic and bosonic particles corresponding to the Chandrasekhar mass for a DM particle of mass $m_\chi$ are given, respectively, by:
\begin{gather}
    N^{\rm fermionic}_{\rm Chandra} \simeq 1.8 \times 10^{57} \left(\frac{ {\rm GeV}}{m_\chi}\right)^3, \\
    N^{\rm bosonic}_{\rm Chandra} \simeq 1.5 \times 10^{38} \left(\frac{{\rm GeV}}{m_\chi}\right)^2.
\label{eq:Nchandra}
\end{gather}

We also note that due to the dispersion velocities of DM, the NS gravity alone cannot capture DM particles, and some dissipative mechanism involving the interaction of DM with hadrons would be needed. For the species of DM considered in this paper, DM particles are assumed to interact with hadrons through the weak interaction.
However, the accumulation of DM particles over the lifetime of the NS for the accretion rate considered in this paper will not significantly increase its mass. 

\subsection{Dark matter capture by neutron stars}
\label{sec:DMcapture}

The ambient DM attracted by the NS's gravity is captured if its trajectory intersects the star, and it loses energy through its interactions with baryons and leptons. The capture rate of DM by gravitating bodies such as the earth and the sun was first computed in~\cite{Press:1985ug} and a general analytic theory was developed in~\cite{Gould:1987ir}. These calculations have been refined to include the effects of matter degeneracy in estimating the capture rates of NSs~\cite{McDermott:2011jp,Garani:2018kkd}. Most recently, the calculations have been further improved to include general relativistic corrections and provide a consistent treatment of several different operator structures that define the interaction between nucleons and DM~\cite{Bell:2020jou}.   

The capture rate, in general, depends on the ambient energy density of DM, $\rho_\chi$; the DM scattering cross-section off targets in the NSs, $\sigma$; the number density of targets, $n_{\rm t}(r)$; the escape velocity in the NS at radius $r$, $v(r)$; the DM velocity dispersion far away from the star, $v_\chi$; and the velocity of the NS, $v_*$. In what follows, we adopt the result derived in~\cite{Bell:2020jou} to obtain the total capture rate for DM in the mass range $1-10^6$ GeV and for a constant cross-section $\sigma$. In this mass range, Pauli blocking suppression of scattering is unimportant and the total capture rate is
\begin{equation}
    C_{\rm NS} = C_{\rm geom}~{\rm Min} \left[\frac{\sigma}{\sigma_{\rm th}},1 \right] \,, \
\end{equation}     
where 
\begin{equation}
C_{\rm geom}= \pi R_*^2\frac{\rho_\chi}{m_\chi} v_*~{\rm Erf} \left[\sqrt{\frac{3}{2}}\frac{v_*}{v_\chi}\right] \frac{(v^2_{\rm esc}/v_*^2)}{1-(v^2_{\rm esc}/c^2)}
\end{equation}
is the geometric capture rate that includes the effect of gravitational focusing in general relativity, and 
\begin{equation}
\sigma_{\rm th} = \frac{\pi R_*^2}{N_{\rm t}~\xi}
\end{equation}
is the threshold value of the cross-section required to ensure that the DM particles traversing the NS have at least one collision. Here, $N_{\rm t}$ is the total number of target particles in the NS, and 
\begin{equation}
\xi= \frac{4\pi}{N_{\rm t}}\int_0^{R_*}  r^2 dr \sqrt{g_{rr}(t)}n_{\rm t}(r)\frac{1-g_{tt}(r)}{1-g_{tt}(R_*)} \frac{g_{tt}(R_*)}{g_{tt}(r)} \,,
\end{equation}
where $g_{tt}$ and $g_{rr}$ are temporal and spatial components of the Schwarzschild metric. The dimensionless number $\xi\approx {\cal O}(1)$, and its precise value depends on the structure of the NS and the  density profile of targets in the NS interior.  

For a typical NS with mass $M_*\simeq 1.4 ~M_\odot$, radius $R_*=12$ km, and total baryon number $N_t\simeq 2 \times 10^{57}$, assuming that $\xi \approx 1$, the threshold cross-section $\sigma_{\rm th} \approx 2 \times 10^{-45} ~{\rm cm}^2$. For this case, the accumulation of DM in a NS at rest, i.e., $v_*=0$, is given by 
\begin{equation}
N_\chi(t) =  \sqrt{6 \pi} R_*^2\frac{\rho_\chi}{m_\chi} \frac{(v^2_{\rm esc}/v_\chi)}{1-(v^2_{\rm esc}/c^2)}
{\rm Min} \left[\frac{\sigma}{\sigma_{\rm th}},1 \right]~t \,.
\end{equation}
Using $v_{\rm esc}=\sqrt{2 GM_*/R_*}=0.6c$, and a DM velocity dispersion $v_\chi=220$ km/s, we obtain 
\begin{equation}
N_\chi(t) = 2.9\times 10^{43}  \frac{\rho_\chi}{\rho_0} \left(\frac{{\rm GeV}}{m_\chi}\right)
{\rm Min} \left[\frac{\sigma}{\sigma_{\rm th}},1 \right]~\frac{t}{10^{10}~{\rm yr}} \,, 
\end{equation}
where $\rho_0=1$ GeV/cm$^3$ is our choice for the fiducial DM density.   

\subsection{Thermalization and formation of a\\mini-black hole}
 The captured DM particles continue to scatter and eventually thermalize with NS matter. A general analysis of thermalization is challenging because the DM energy decreases over several orders of magnitude during this process. At low energy, matter degeneracy and correlations due to strong interactions between baryons can substantially alter the scattering rate~\cite{Bertoni:2013bsa,Garani:2020wge}. For  scattering off neutrons with a  constant cross-section, the thermalization time for the DM masses of interest is
\begin{equation}
        t_{\rm th} \simeq 3750\ \text{years} 
        \frac{\gamma}{(1+\gamma)^2}
        \left(\frac{2 \times 10^{-45} \rm {cm^2}}{\sigma_{\rm \chi}}\right) 
        \left(\frac{10^5\rm K}{\rm T}\right)\,,
\end{equation}
where $\gamma = m_{\chi}/m_{\rm n}$, and $m_{\rm n}$ is the neutron mass~\cite{Bertoni:2013bsa}. 

Upon thermalization, the radius of the DM sphere is determined by the temperature and the gravitational potential of the NS~\cite{Bertoni:2013bsa}
\begin{equation}
        r_{\rm{th}} \approx 2.2\ {\rm m} \left(\frac{{\rm T}}{10^5\ {\rm K}}\right)^{1/2} \left(\frac{{1~\rm GeV}}{m_\chi}\right)^{1/2}\,.
\end{equation}

When the mass density of bosonic DM exceeds that of baryons, the DM becomes self-gravitating. This occurs when the total number of DM particles exceeds
\begin{equation}
        N_{\rm {self}} \simeq 4.8 \times 10^{46} 
        \left(\frac{\rm{GeV}}{m_\chi}\right)^{5/2}
        \left(\frac{\rm{T}}{10^5\ \rm{K}}\right)^{3/2}\,. 
\label{eq:Nself}
\end{equation}
The self-gravitating DM sphere collapses to form a BH when its mass exceeds the Chandrasekhar limit for bosonic matter, i.e., if $N_{\rm {self}} > N_{\rm Chandra}^{\rm bosonic}$. Using Eq.~\eqref{eq:Nchandra}, we can deduce that $N_{\rm self} > N_{\rm Chandra}^{\rm bosonic}$ for $m_\chi \leq 10^{17}~\rm GeV(T/10^5~\rm K)^3$. Thus, for the mass range $m_\chi \in [1,\,10^6]$~GeV considered here, a mini-BH of mass 
\begin{equation}
M_{\rm BH}= m_\chi  N_{\rm {self}} \simeq 4.8 \times 10^{46} 
        \left(\frac{\rm{GeV}}{m_\chi}\right)^{3/2}
        \left(\frac{\rm{T}}{10^5\ \rm{K}}\right)^{3/2}{\rm GeV}
\label{eq:M_BH}
\end{equation} 
forms at time 
\begin{equation}
\begin{split}
    t_{\rm BH} &= \frac{N_{\rm {self}}}{C_{\rm geom}} {\rm Max} \left[\frac{\sigma_{\rm th}}{\sigma},1 \right] \\
    &\approx 10^{13}~\left(\frac{\rho_0}{\rho_\chi}\right) \left(\frac{\rm GeV}{m_\chi}\right)^{3/2} {\rm Max} \left[\frac{\sigma_{\rm th}}{\sigma},1 \right]~{\rm yr}
\end{split}
\label{eq:tc accretion}
\end{equation} 
when $N_\chi(t_{\rm BH})=N_{\rm self}$. For $\rho_\chi \simeq \rho_0$, collapse occurs at $t<10^{10}$~yrs for $m_\chi\gtrsim 100$~GeV. 

\subsection{Bose-Einstein Condensation} 
 In the preceding discussion, we neglected the formation of a Bose-Einstein Condensate (BEC) of DM. Earlier work has shown that a BEC can accelerate BH formation for bosonic DM with mass $m_\chi <   m_{\rm BEC}$, where $m_{\rm BEC} \simeq 2 \times 10^4$ GeV for $T=10^5$ K and $m_{\rm BEC} \simeq 5 \times 10^3$ GeV for $T=10^6$ K~\cite{Kouvaris:2011fi,McDermott:2011jp}. We briefly summarize their findings below.  
 
When the critical temperature for BEC of DM     

\begin{equation}
T_C = \frac{2\pi}{m_\chi}\left( \frac{3 N_\chi(t)}{4\pi r_{\rm th}^3~\zeta(3/2)} \right)^{2/3}
\end{equation}
exceeds the NS temperature, a condensate forms at the core within a radius
\begin{equation}
 r_{\rm BEC} = 1.5 \times 10^{-4} {\rm cm}\left(\frac{1{\rm GeV}}{m_\chi}\right)^{1/2}
\end{equation}
which is much smaller than $r_{\rm th}$. This occurs when the total number of accumulated DM,
\begin{equation}
N_\chi(t)= N_{\rm C}=10^{36}~\left( \frac{T}{10^5~{\rm K}}\right)^3\,,    
\end{equation}
and the number of DM particles in the condensate,
\begin{equation}
N_{\rm BEC}(t)= N_\chi(t)- N_\chi(t) \left(\frac{T}{T_C}\right)^{3/2}= N_\chi(t) - N_C\,,    
\end{equation}
increase at the accretion rate on timescales that are large compared to the thermalization time. The BEC becomes self-gravitating when $N_{\rm BEC} > 4\pi r_{\rm BEC}^3 \rho_B/m_\chi \approx 10^{28} ({\rm GeV}/m_\chi)$. Since  $N_{\rm BEC} \ll N^{\rm  bosonic}_{\rm Chandra}$ for the mass range of interest, the self-gravitating BEC is stable, and the time required for it to collapse is obtained as a solution to $N_\chi( t^{\rm BEC}_{\rm BH}) = N^{\rm  bosonic}_{\rm Chandra} + N_C$. We find that a BH of mass 
\begin{equation} 
M^{\rm BEC}_{\rm BH} = m_\chi N^{\rm  bosonic}_{\rm Chandra} \simeq 1.5\times 10^{38} \left(\frac{\rm GeV}{m_\chi} \right)\rm GeV
\label{eq:M_BH_BEC}
\end{equation}  
forms at a time 
\begin{equation}
\begin{split}
          t^{\rm BEC}_{\rm BH} &= \frac{N^{\rm  bosonic}_{\rm Chandra} + N_C}{C_{\rm geom}} {\rm Max} \left[\frac{\sigma_{\rm th}}{\sigma},1 \right]\\
          &\approx 5.2\times 10^4 \left(\frac{\rho_0}{\rho_\chi}\right) \left(\frac{\rm GeV}{m_\chi}\right) {\rm Max} \left[\frac{\sigma_{\rm th}}{\sigma},1 \right] \\
          & \times \left[ 1+ \frac{2}{3} \left(\frac{m_\chi}{10 ~\rm GeV}\right)^2 \left(\frac{T}{10^5~{\rm K}}\right)^3 \right] {\rm yr}\,.
        \end{split}
        \label{eq:tc BEC}
\end{equation} 
At $T=10^5$~K, this equation implies $t^{\rm BEC}_{\rm BH} < t_{\rm BH}$ for $m_\chi <  2 \times 10^4$~GeV, and BEC greatly reduces the time for BH formation in this mass range. For example, when $m_\chi  = 1$~GeV, the BEC reduces the collapse time by $\approx 10^8$~yrs. Consequently, mini-BH formation is possible within the Universe's lifetime for the range of DM considered in this study ($m_\chi  = 1-10^6$ GeV). 

\subsection{Growth of the black hole due to accretion of baryons}

The mini-BH continues to grow if the Bondi-Hoyle accretion rate, 
$\dot{M}_{\rm{acc}} = 4 \pi \lambda_s (GM_{\rm{BH}}/v_{\rm{s}}^2)^2\rho_{\rm{C}} v_{\rm{s}}$, exceeds the Hawking evaporation rate, $\dot{M}_{\rm{evap}} = (15360 \pi G^2 M^2_{\rm{BH}})^{-1}$, where $v_s$ is the sound speed, $\rho_{\rm C}$ is the central baryon density of the NS surrounding the BH, $M_{\rm BH}$ is the mass of the BH at the star's core, and $\lambda_s$ is a dimensionless constant that depends on the EOS of matter in the NS core. 
The growth is possible only if the BH reaches a critical mass, 
\begin{equation}
M_{\rm{BH}}^{\rm{crit}} \simeq 1.3 \times 10^{37} \left(\frac{v_s}{0.3~c}\right)^{3/4} \left(\frac{10^{15} {\rm g/cm}^3}{\lambda_s \rho_B}\right)^{1/4} {\rm GeV}\,. 
\label{eq:M_BH_C}
\end{equation} 
The analysis of BH growth rates presented in Refs.~\cite{McDermott:2011jp,Kouvaris:2011fi} use $\lambda_s=1/4$, corresponding to a polytropic index $\Gamma=5/3$. Since the EOS of the NS core is expected to be stiff, with $\Gamma \simeq 2$ and $v_s \simeq c/\sqrt{3}$, there has been much recent work on understanding Bondi-Hoyle accretion under these conditions~\cite{East:2019dxt,Baumgarte:2021thx,Richards:2021upu,Richards:2021zbr,Aguayo-Ortiz:2021jzv,Giffin:2021kgb}. These studies suggest $\lambda_s\simeq 1.3$ for realistic NS EOSs.  

\begin{figure*}
     \centering
     \includegraphics[width=\textwidth]{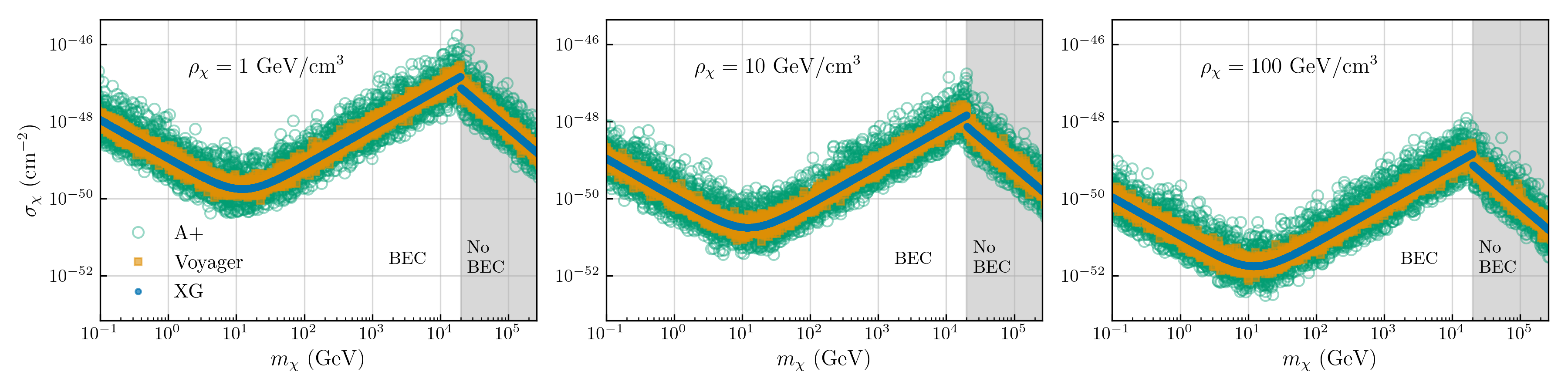}
    \caption{Distribution for the mass $m_{\chi} \in [10^{-1}$, $2.5\times10^5]\ {\rm GeV}$ and scattering cross-section $\sigma_{\chi}$ inferred from the collapse time derived from the observed number of BBHs by the detector networks considered in Fig.~\ref{fig:NBBH and tcollapse}. Here we consider the scenario where the NS implodes without the formation of a BEC state for $m_\chi \geq 2 \times 10^4\ {\rm GeV}$, shown in grey. For $m_\chi < 2 \times 10^4\ {\rm GeV}$, NS implosion occurs at shorter timescales through the formation of a BEC. We use $\rho_{\chi}=[1, 10, 100]\ {\rm GeV/cm^3}$.}
    \label{fig:dm-posterior}
\end{figure*}

For the scenario in which a BH forms without a BEC intermediate state, i.e., for $m_\chi \gtrsim m_{\rm BEC}$, from Eq.~\eqref{eq:M_BH} and Eq.~\eqref{eq:M_BH_C} we find that $M_{\rm BH}> M_{\rm{BH}}^{\rm{crit}}$ for the mass range of interest. In the scenario that involves a BEC intermediate state,  i.e., for $m_\chi < m_{\rm BEC}$, comparing Eq.~\eqref{eq:M_BH_BEC} and Eq.~\eqref{eq:M_BH_C} we find that $M_{\rm BH}> M_{\rm{BH}}^{\rm{crit}}$ is only satisfied for $m_\chi \lesssim 10$~GeV. For larger $m_\chi$ the BH evaporates, because the accretion of baryons cannot keep pace with mass loss from Hawking radiation. 

\begin{figure}[t]
    \includegraphics[width=0.99\columnwidth]{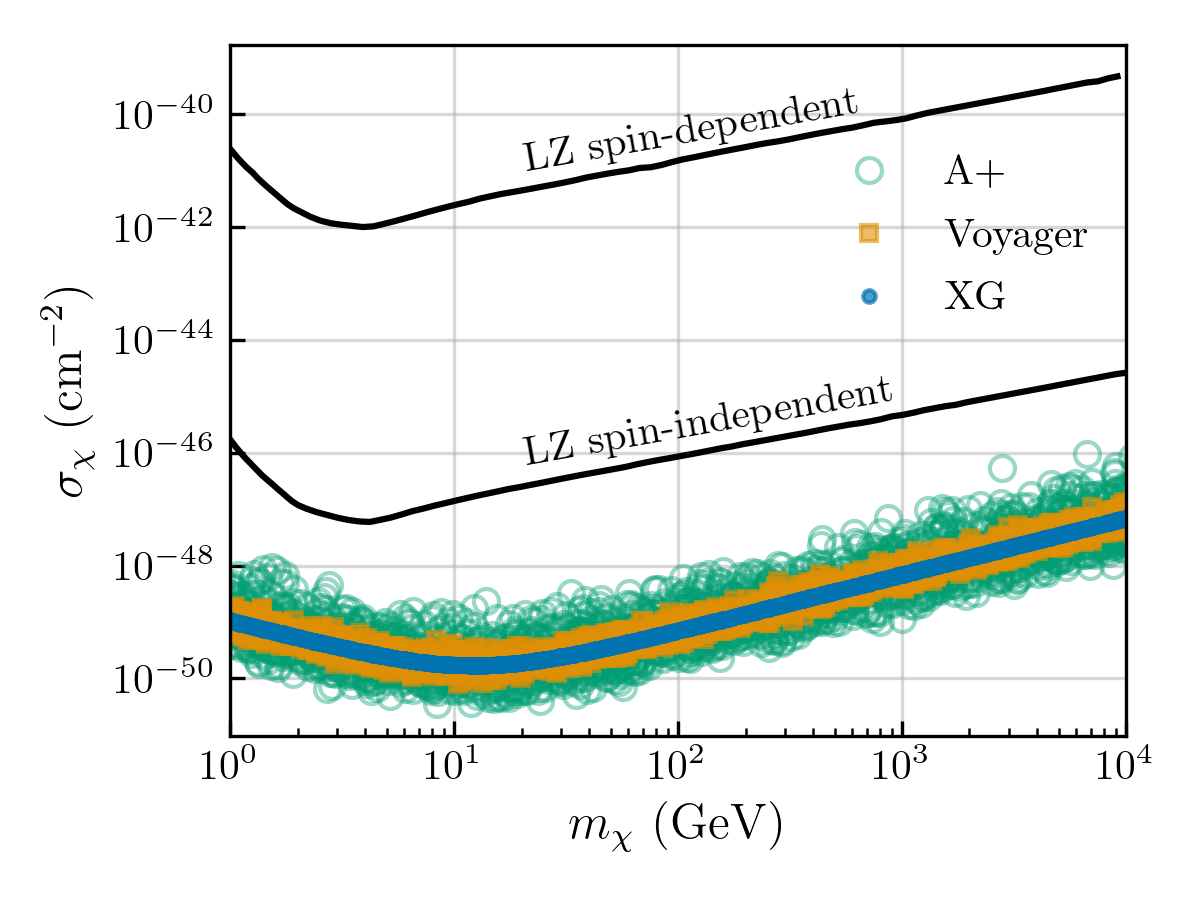}
    \caption{Comparison between the constraints obtained for $m_\chi \in [1, 10^4]$ GeV with the lowest ambient DM density considered in this work ($\rho_\chi = 1\ {\rm Gev/cm^3}$) with the latest constraints from the direct detection experiment, LZ~\cite{LZ:2022ufs}.}
    \label{fig:comparison with LZ}
\end{figure}

\subsection{Growth due to dark matter accretion}  
If the thermalization time is short compared to the BH Hawking evaporation time $t_{\rm Haw} = 15360 \pi G^2 M^3_{BH}/3 \simeq 5 \times 10^{4} (100~{\rm GeV}/m_\chi)^3$, DM accreting onto the NS reaches the BEC efficiently. The maximal impact parameter for DM capture is larger than the radius of the BEC, and DM particles reaching the BEC can feed the growth of the BH at a rate $\dot{M}_{\rm DM} = C_{\rm NS} t$ if Hawking radiation does not disrupt the BEC~\cite{McDermott:2011jp}. Direct heating of DM will be absent if the dark sector does not contain mediator particles with mass less than the Hawking temperature  $T_{\rm haw}=1/(8\pi G M_{\rm BH})$. Further, even when DM is directly heated by Hawking radiation, the change in its temperature would be negligible if it can thermalize with baryons quickly~\cite{McDermott:2011jp}. In this case, $\dot{M}_{\rm DM} = C_{\rm NS} t > \dot{M}_{\rm{evap}}$, and the accretion of DM prevents BH evaporation for heavier masses~\cite{McDermott:2011jp}. Comparing the timescales for thermalization and evaporation, we find that DM accretion can prevent BH evaporation for $m_\chi \lesssim 4 \times 10^3~(\sigma_{\chi}/\sigma_{\rm th}) (T/10^5 ~{\rm K})$~GeV.

\subsection{Neutron star implosion timescale}
The timescale for a mini-BH to devour the entire NS by Bondi-Hoyle accretion has been computed recently for the stiff EOS expected in NS cores~\cite{Baumgarte:2021thx}. For a realistic NS model, they estimate this timescale to be given by    
\begin{equation}
t_{\rm dev} \simeq 0.3  \left(\frac{10^{46}~{\rm GeV}}{M_{\rm BH}}\right){\rm yr}\,,      
\end{equation}
where $M_{\rm BH}$ is the mass of the initial mini-BH. For large DM mass, i.e., $m_\chi> m_{\rm BEC}$ where $m_{\rm BEC} \simeq 2 \times 10^4$ GeV at $T=10^5$ K, the NS is destroyed on a timescale $t \approx 6\times 10^{-2} (m_\chi/{\rm GeV})^{3/2} (10^5~{\rm K}/T)^{3/2}~{\rm yr}$. For $m_\chi < m_{\rm BEC}$ (which involves a BEC intermediate state), the NS is destroyed on a timescale $t \approx 3.2 \times 10^7 (m_\chi/{\rm GeV})~{\rm yr}$.

\section{Constraining dark matter properties from the implosion timescale}
\label{sec:constraints}
As discussed before, GW observations can determine the relative abundance of BBHs and BNSs, which allows us to infer the implosion timescale $t_c.$ The posterior distribution of $t_c$, or the lower bound on $t_c$ if GW observations can not conclusively measure a nonzero BBH rate, can be used to constrain a region in the space of DM mass, interaction cross section and dispersion velocity in the vicinity of merger.  To illustrate the sensitivity of this method, we assume that the time it takes for DM particle capture to form a BH of critical mass, $M^{\rm{crit}}_{\rm{BH}}$, is approximately equal to the total collapse time $t_c$ of the NS to form a BH through this channel. Furthermore, for illustration, we assume that $t_c=10\,\rm Gyr.$

Figure~\ref{fig:dm-posterior} shows the constraint on the DM particle mass, $m_{\chi}$, and the DM-baryon interaction cross-section, $\sigma_{\chi}$, for the inferred collapse times shown in Fig.~\ref{fig:NBBH and tcollapse} and three values of the DM density: $\rho_{\chi}=1\,{\rm GeV/m^3},$ ${\rm 10\, GeV/m^3}$, and $100\,{\rm GeV/m^3}.$ In all cases the dispersion velocity is assumed to be 200 km/s. The precision of the inference of DM parameters is directly related to how precisely we can deduce the collapse time from the observed number of BBH mergers and the measured $\tilde{\Lambda}$ from the GW signal. Compared to the A+ network, the XG network of Cosmic Explorer and Einstein Telescope can improve the width of the distribution by almost a factor of 10. From Fig.~\ref{fig:comparison with LZ}, these constraints are competitive with those of direct detection DM experiments, such as the LUX-ZEPLIN (LZ) experiment~\cite{LZ:2022ufs}, over the entire range of $m_\chi$.

\section{Conclusions and Future prospects}
\label{sec:conclusions}

In the last decade, direct detection experiments for DM have proven to be successful in constraining the parameter space for WIMPs as well as axionic DM~\cite{LZ:2022ufs, DarkSide-50:2022qzh, XENONCollaboration:2022kmb, SENSEI:2020dpa, XENON:2018voc}. Microlensing surveys~\cite{Macho:2000nvd,EROS-2:2006ryy}, calculations from dwarf-galaxy dynamics~\cite{2011MNRAS.416.2949W,Koushiappas:2017chw}, as well as searches for sub-solar mass compact binary mergers in GW data~\cite{LIGOScientific:2005fbz,LIGOScientific:2018glc,LIGOScientific:2019kan,LIGOScientific:2021job} have probed the DM compact object parameter space, deriving limits on the abundance of DM in these objects and on their mass spectrum.

In this work, we present a method to combine measurements from GW observations with the particle properties of DM, especially in the WIMP mass range, through the observation (or lack thereof) of a novel population of BBHs in the mass range $1M_\odot \leq m \leq 2 M_\odot$. With XG ground-based GW detectors, our ability to measure the effective tidal deformability will improve tremendously, as shown in Fig.~\ref{fig:lambda cdf}. With this refinement, we expect to identify a significant number of BBH mergers in this mass range, if this population exists in the Universe. The observed number of mergers can potentially constrain the implosion time of NSs if these BHs form from the collapse of old NSs due to the presence of a mini-BH at their cores. If the mini-BH forms through the accumulation of DM in the cores of NSs over their lifetime, the collapse time can then inform us on DM particle properties. 

We illustrate how this mechanism works. We report limits on the interaction cross-section of DM particles with baryons and particle masses for a simplistic scenario where asymmetric DM interacts weakly with hadrons and gets captured through scattering within NS cores to form a self-gravitating mass that forms a BH. As expected, the constraints are more precise in the case of XG GW detectors, which is a direct consequence of how precisely the binary's effective tidal deformability, and hence the collapse time, is inferred from the observed number of BBH mergers. 

One potential issue is the degeneracy with primordial BHs, which could also populate this spectrum of masses. One possible way to remove the degeneracy involves accurate spin measurements to distinguish between the two formation channels. Furthermore, if primordial BHs contribute to the super-solar mass range, we would also expect to observe some sub-solar mass primordial BHs. The absence of observed old NSs also complements the observation of BHs formed from imploding NSs. In this work, we show how effectively GW detections can not only constrain DM interactions, but also rule out models that allow for NSs to implode through DM accumulation if no such population is observed in the future.

\section*{Acknowledgments}

We thank K.~Belczynski for useful discussions on BNS delay times, as well as M.~Baryakhtar and T.~Slatyer for discussions on the DM scenarios considered in this paper.
D.S. and B.S.S. were supported in part by NSF grant No.\,PHY-1836779, PHY-2012083, AST-2006384 and PHY-2207638. A.G. is supported by NSF grant No. AST-2205920.
E.B. is supported by NSF Grants No. AST-2006538, PHY-2207502, PHY-090003 and PHY20043, and NASA Grants No. 19-ATP19-0051, 20-LPS20-0011 and 21-ATP21-0010. This research project was conducted using computational resources at the Maryland Advanced Research Computing Center (MARCC).
S.R. is supported by the U.S. Department of Energy under Grant No. DE-FG02- 00ER41132 and National Science Foundation’s Physics Frontier Center: The Network for Neutrinos, Nuclear Astrophysics, and Symmetries.  This paper has the INT preprint number INT-PUB-22-028.
Part of E.B.'s and B.S.S.'s work was performed at the Aspen Center for Physics, which is supported by National Science Foundation grant PHY-1607611. This research was also supported in part by the National Science Foundation under Grant No. NSF PHY-1748958.

\bibliographystyle{apsrev4-1}
\bibliography{implodingns}

\end{document}